\documentclass[aps,preprintnumbers,superscriptaddress,onecolumn,amsmath,amssymb,floatfix,10pt,pra]{revtex4-1} 
\usepackage{graphicx} 

\usepackage[colorlinks,bookmarks=false,citecolor=violet,linkcolor=red,urlcolor=blue]{hyperref}
\usepackage{xcolor}
\usepackage{amsfonts}
\usepackage{amsmath}
\usepackage{amssymb}
\usepackage{amsthm}

\usepackage{bm}
\usepackage{physics}
\usepackage{slashed}

\usepackage{mathtools}
\usepackage{tikz}
\usepackage{braket}
\usepackage{verbatim}
\newcommand{\eps}{\varepsilon}

\newcommand{\ci}{\mathrm{i}\mkern1mu}

\begin{document}
\title{Higher-order tacnodes in a free fermionic model}

\author{Andrea Maroncelli}
\affiliation{Dipartimento di Fisica e Astronomia, Università di Firenze, and INFN, Sezione di Firenze, Via G. Sansone 1, I-50019 Sesto Fiorentino (FI), Italy}
\email{andrea.maroncelli@unifi.it}
\author{Jean-Marie St\'ephan}
\affiliation{CNRS, ENS de Lyon, LPENSL, UMR5672, 69342, Lyon cedex 07, France}
\email{jean-marie.stephan@ens-lyon.fr}

   \begin{abstract}
We investigate a one-dimensional free fermion model with nearest and next-nearest neighbor hopping, evolving in imaginary time from a product state with $N$ consecutive fermions, and conditioned to go back to the same state after a given time. Such types of models are quantum reformulations of well-studied two-dimensional classical lattice models, which are known to give rise to limit shapes where expectation values of simple local observables, such as density, depend on position in an appropriate scaling limit.
In the case of only nearest neighbor hopping, this model is known to have two fluctuating regions which can be tuned to merge depending on ratio between time and $N$. Correlations near the merger are governed by a so-called tacnode kernel. Here we show that another universal higher-order tacnode process can appear upon including the next-nearest neighbor term. We also discuss the limit shapes, and compute analytically the corresponding density profile.

\end{abstract}

\maketitle
\tableofcontents
\newpage
\section{Introduction}
The study of classical statistical models with boundaries that strongly break translation invariance naturally leads to the phenomenon of \emph{limit shapes}, that is the appearance of a macroscopic state with some inhomogeneous density profile, say of magnetization or particle density. {Typically there are also distinct ordered (or frozen) and disordered (or fluctuating) regions sharply separated by some curves, called \emph{arctic curves}}. Perhaps the most famous such curve is the ``arctic circle'' which was studied in \cite{ArcticCircle} for dimers or vertex models. This phenomenon arises in the scaling limit, where the overall system size is fixed while the number of lattice sites diverges and the lattice spacing tends to zero.
As a result, the total free energy of the system can and does depend on boundary conditions. 

A related example will be provided and explained in Figure \ref{fig:limitshapes_alpha0}, where the frozen regions are shown in deep blue (vanishing density) and yellow (density one). For interacting systems determining such arctic curves is a daunting task, despite some achievements accumulated over the years \cite{ColomoPronkocurve,ColomoPronkoZinn,deGier2021}. In contrast, for free systems a general framework based on a variational (minimization) principle \cite{Nienhuis_1984,variationaldimers} makes it possible to characterize these limit shapes and their arctic curves, at least in an implicit form.

Even if many were neither formulated nor studied as such, almost all known limit shapes problems which have been investigated in this context can be reformulated as spins or fermionic models evolving in discrete or continuous imaginary time (e.g. \cite{PrahoferSpohn,Allegra_2016,Pallister_2022}), thanks to the well-known transfer matrix (Hamiltonian) formalism. A crucial role is played by the $U(1)$ symmetry of the transfer matrix: this symmetry implies that particles cannot rearrange easily to reach a simple homogeneous ground state situation through imaginary time evolution, as would be the case say for the Ising model.

Moreover, the behavior of correlations near the edge of the fluctuating region (close to the arctic curve) has attracted considerable interest due to its universal behavior. The most common case is governed by the Pokrovsky-Talapov scaling law \cite{PokrovskyTalapov} with exponent $1/3$: non-trivial fluctuations occur on a scale $R^{1/3}$, where $R$ denotes the typical size of the fluctuating region. The rescaled correlations have been studied extensively in the mathematical literature under the name \emph{Airy process}, which was introduced originally in \cite{PrahoferSpohn}. This universal edge behavior extends well beyond the context of limit shapes, appearing in a wide range of seemingly unrelated problems, including the edge spectrum of random matrices \cite{Jurkiewicz1983,Periwal,TracyWidom1994}, random Young tableaux \cite{kerov_vershik}, crystal surfaces \cite{PokrovskyTalapov,bcc_crystals}, quantum quenches \cite{EislerRacz2013}, ground states of fermionic gases \cite{Eislertrapped,Dean_2019} in trapping potentials, or spin chains in slowly varying magnetic field, classical exclusion processes \cite{TracyWidom2009}, to name only a few. There are also known relations to the KPZ universality class \cite{KPZ,Takeuchi1,Takeuchi2}.

Other edge universality classes are possible as well.
An important feature of all those behaviors lies in the fact that, unlike bulk correlations, they are not modified by the effects of (sufficiently local) interactions. This is because the particle or hole density is small near the edge, and interactions can safely be neglected, see e.g. \cite{Stephan_free}. So in essence we are always brought back to some form of free fermions when studying the edge. However, depending on certain other features of the model (such as dispersion at low momenta of the underlying free fermions), or the geometric nature of the arctic curves, one can obtain new universality classes, with the same or different scaling laws. For example, much effort has gone into studying other processes such as the Pearcey process which appears near a cusp. Higher Airy processes \cite{LeDoussalMajumdarSchehr2018} occur for quartic or higher dispersion at low momenta \cite{LeDoussalMajumdarSchehr2018} contrary to the usual quadratic one, and those can also appear in ground states of one-dimensional fermions in traps, and they come with different exponents. Another class is provided by the so-called \emph{tacnode processes} \cite{2011_Delvaux,Adler2013,Johansson2013,Adler2014}. The name tacnode comes from algebraic geometry: the simplest tacnode occurs when two parabolas touch, corresponding to the algebraic curve $y^{2}=x^{4}$ (or, in our case, $x^{2}=y^{4}$). Such a situation can be observed locally in the middle of Figure \ref{fig:limitshapes_alpha0}(a), where the two arctic circles merge, and for this reason it was called \emph{merger transition} in \cite{Pallister_2022}. The exponent is the same as for Airy, but the underlying process is much more complicated, due to the fact that the two edges subtly influence each other.

In this paper, we investigate a new edge universality class, which naturally occurs with both a tacnode geometry and a higher than quadratic dispersion at low momenta. In particular, we study the \emph{simplest microscopic model} in which such new universality class occurs and can be studied fully analytically. This is a tight-binding free fermion Hamiltonian on the line with an extra next-nearest neighbor hopping term studied in \cite{Bocini2021}. However, we consider its evolution from a different initial condition: a ``double domain-wall'' state, in which $N$ particles occupy consecutive sites at some position along the line. This model enables us to combine the aforementioned ingredients necessary to engineer a tacnode situation where two fluctuating regions nearly touch, depending on both size $N$ and imaginary time duration $R$. By suitably tuning the next-nearest neighbor coupling we are also able to get quartic behavior for the fermions and to get higher-order behavior, similar to \cite{Stephan_free,Walsh2023}. The corresponding correlations are governed by a new higher-order tacnode kernel which we investigate. In passing we study the corresponding limit shapes and transitions in the free energy. 

The paper is organized as follows. In Section \ref{sec:introknown} we introduce the model, the imaginary time setup studied here, and discuss known related results on the free energy. In Section \ref{sec:correlations} we study the corresponding limit shapes and density profile. Our results are based on various exact formulas for the fermionic two-point function which are summarized at the beginning. Edge correlations are studied in section \ref{sec:edge}, which culminates with the higher-order tacnode kernel advertised in the introduction. Conclusions and perspectives are summarized in Section \ref{sec:conclusion}. Our results rely heavily on the theory of orthogonal polynomials, and several appendices \ref{app:orthopoly}, \ref{app:asymptotics_Bessel}, \ref{app:freefermions}, \ref{app:besselfunction} provide the necessary background on general such polynomials, the specific weight corresponding to our fermionic problem, and various other free fermions or asymptotic calculations.

\section{The model, free energy, and known results}
\label{sec:introknown}
\subsection{Free fermions in imaginary time}
The model under study is a system of fermions governed by the following tight-binding Hamiltonian:
\begin{equation}\label{eq:Hcoordinate}
	H=\frac{1}{2} \sum_{x \in \mathbb{Z}} \bigg[c_{x+1}^{\dagger}c_x+c^{\dagger}_xc_{x+1}+\alpha \big(c_{x+2}^{\dagger}c_x+c^{\dagger}_xc_{x+2}\big) \bigg],
\end{equation}
where $\alpha\geq 0$ and the $c_x$, $c_x^{\dagger}$ are fermionic operators of annihilation and creation of particles at lattice site $x$, respectively. They satisfy the canonical anti-commutation relations $\{c_x,c_y^{\dagger}\}=\delta_{xy} $, $ \{c_x,c_y\}=\{c_x^{\dagger},c_y^{\dagger}\}=0$. The Hamiltonian \eqref{eq:Hcoordinate} can be rewritten in diagonal form in momentum space
\begin{equation}\label{eq:diagonalH}
	H=\int_{-\pi}^{\pi} \frac{dk}{2\pi} \eps(k)d^{\dagger}(k)d(k),
\end{equation}
where 
\begin{equation}\label{c(k)c(k)+}
		d(k)=\sum_{x\in \mathbb{Z}} e^{-\ci kx} c_x \qquad, \qquad 
        c_x=\int_{-\pi}^{\pi}\frac{dk}{2\pi} e^{\ci k x}d(k)
	\end{equation}
 are the Fourier transform of the operators $c_x$. They satisfy the anti-commutation relation $\{d(k),d^\dag (q)\}=2\pi \delta(k-q)$. The dispersion $\eps(k)$ is given by
\begin{equation}\label{eq:dispersion}
	\eps (k)=\cos(k)+\alpha\cos(2k).
\end{equation}

\begin{figure}[t]
	\centering
	\includegraphics[scale=0.5]{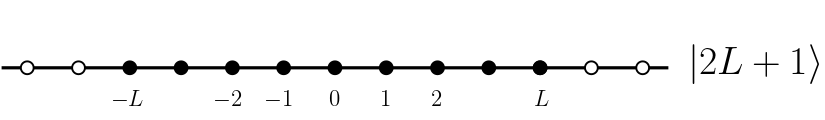}
    \caption{Sketch of the initial and final states of the model $\ket{N}=\ket{2L+1}$, which has fully occupied sites in the interval $[-L,L]$.}
    \label{fig:boundarystate}
\end{figure}

\begin{figure}[htbp]
\begin{tikzpicture}
\begin{scope}
\node at (0,0) {\includegraphics[width=0.45\textwidth]{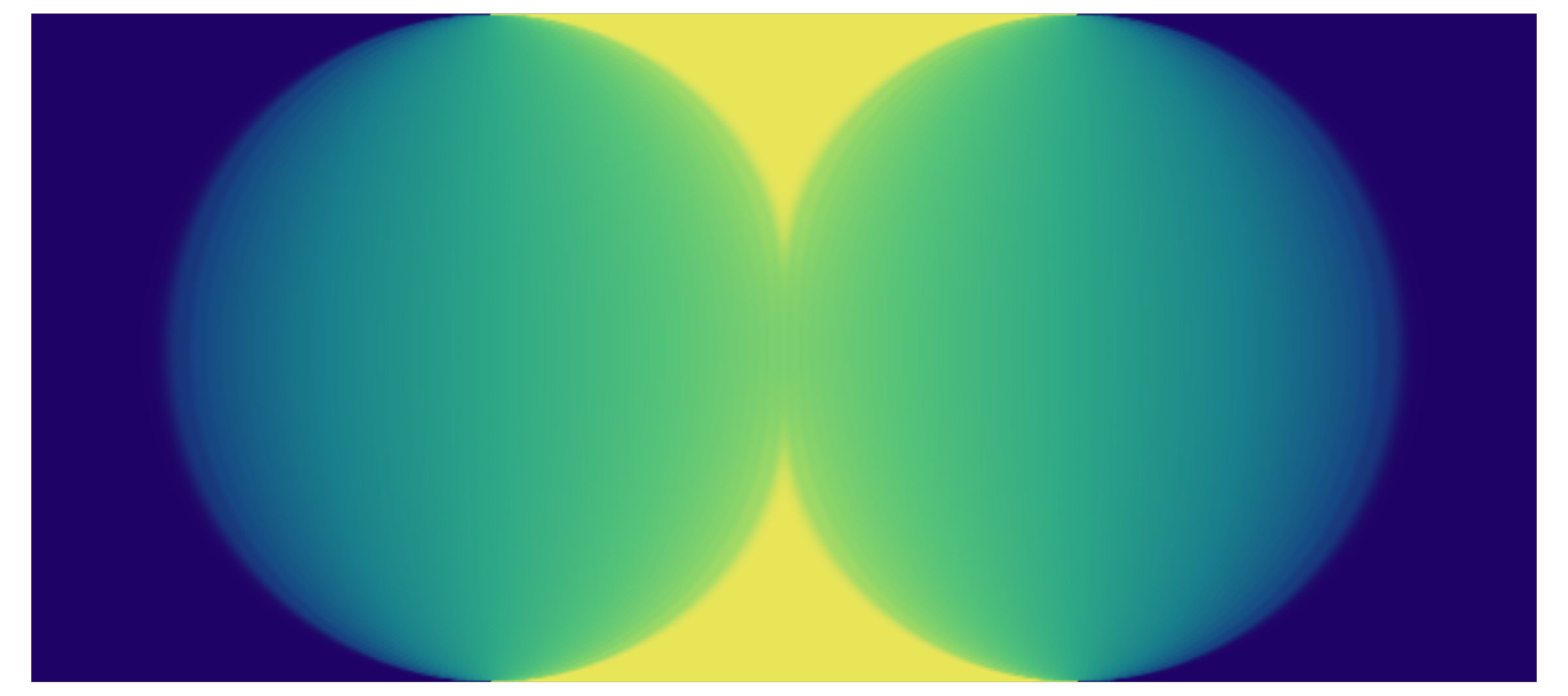}};
\draw[densely dashed,thick] (-4.1,0) -- (4.1,0);

\draw (-2*1.73,-2) node {};
\draw[ultra thick] (-1.6,-1.8) -- (-1.6,-1.6);
\draw (-1.6,-2) node {$-\lambda$};
\draw[ultra thick] (0,-1.8) -- (0,-1.6);
\draw (0,-2) node {$0$};
\draw[ultra thick] (1.6,-1.8) -- (1.6,-1.6);
\draw (1.6,-2) node {$\lambda$};
\draw[ultra thick] (-3.86,-1.7) -- (3.86,-1.7);
\end{scope}
\begin{scope}[xshift=4.8cm,yshift=5cm]
\node at (0,0) {\includegraphics[width=0.45\textwidth]{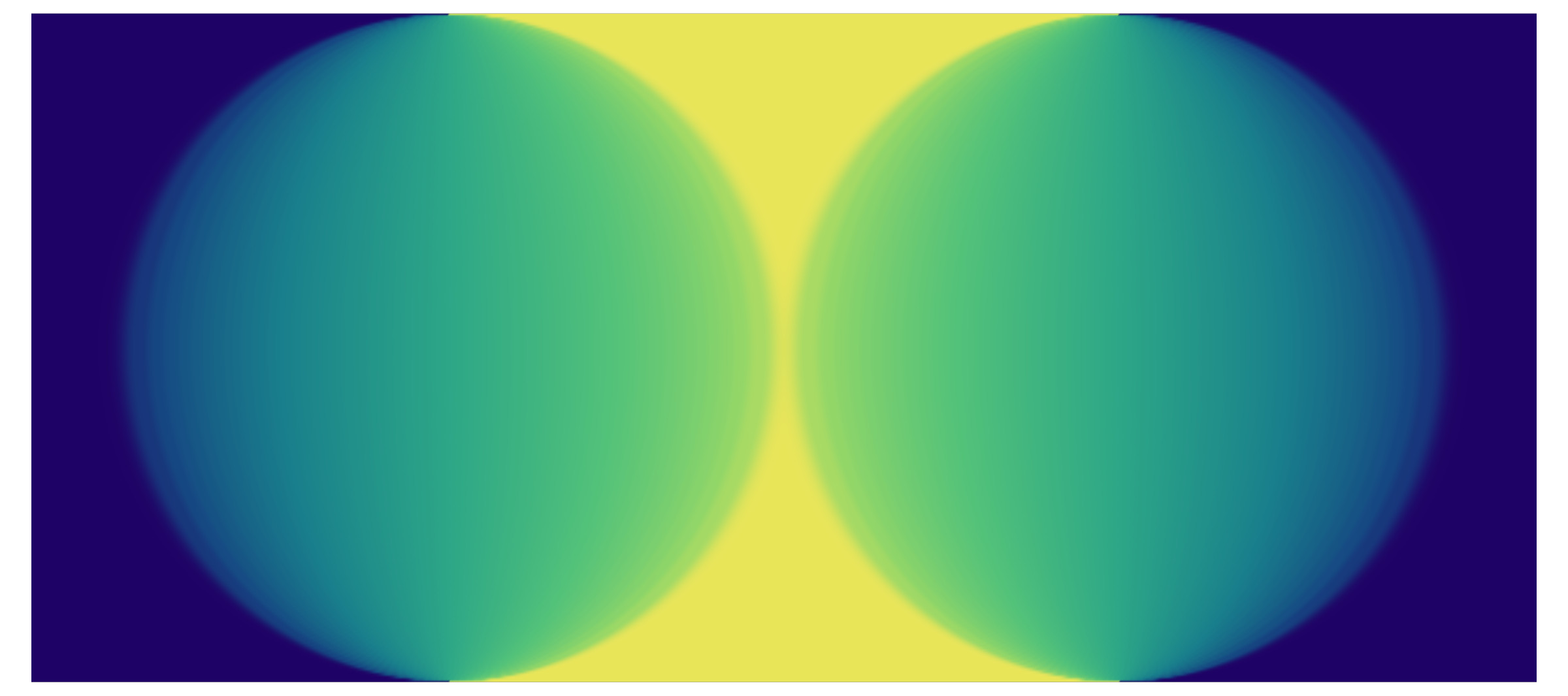}};
\draw (0,2) node {(a)};
\draw[ultra thick] (-3.95,-1.7) -- (-3.8,-1.7);
\draw[ultra thick] (-3.95,1.7) -- (-3.8,1.7);
\draw (-4.2,1.75) node {$\phantom{+}1$};
\draw (-4.2,-1.75) node {$-1$};
\draw[densely dashed,thick] (-4.1,0) -- (4.1,0);
\draw[ultra thick] (0,-1.8) -- (0,-1.6);
\draw (0,-2) node {$0$};
\draw[ultra thick] (-3.86,-1.7) -- (3.86,-1.7);
\draw[ultra thick] (1.72,-1.8) -- (1.72,-1.6);
\draw (1.72,-2) node {$\lambda$};
\draw[ultra thick] (-1.72,-1.8) -- (-1.72,-1.6);
\draw (-1.72,-2) node {$-\lambda$};
\draw (-4.8,-3) node {(b)};
\draw[ultra thick] (-8.75,-3.3) -- (-8.6,-3.3);
\draw[ultra thick] (-8.75,-6.7) -- (-8.6,-6.7);
\draw (-9,-3.25) node {$\phantom{+}1$};
\draw (-9,-6.75) node {$-1$};
\end{scope}
\begin{scope}[xshift=9.6cm]
\node at (0,0) {\includegraphics[width=0.45\textwidth]{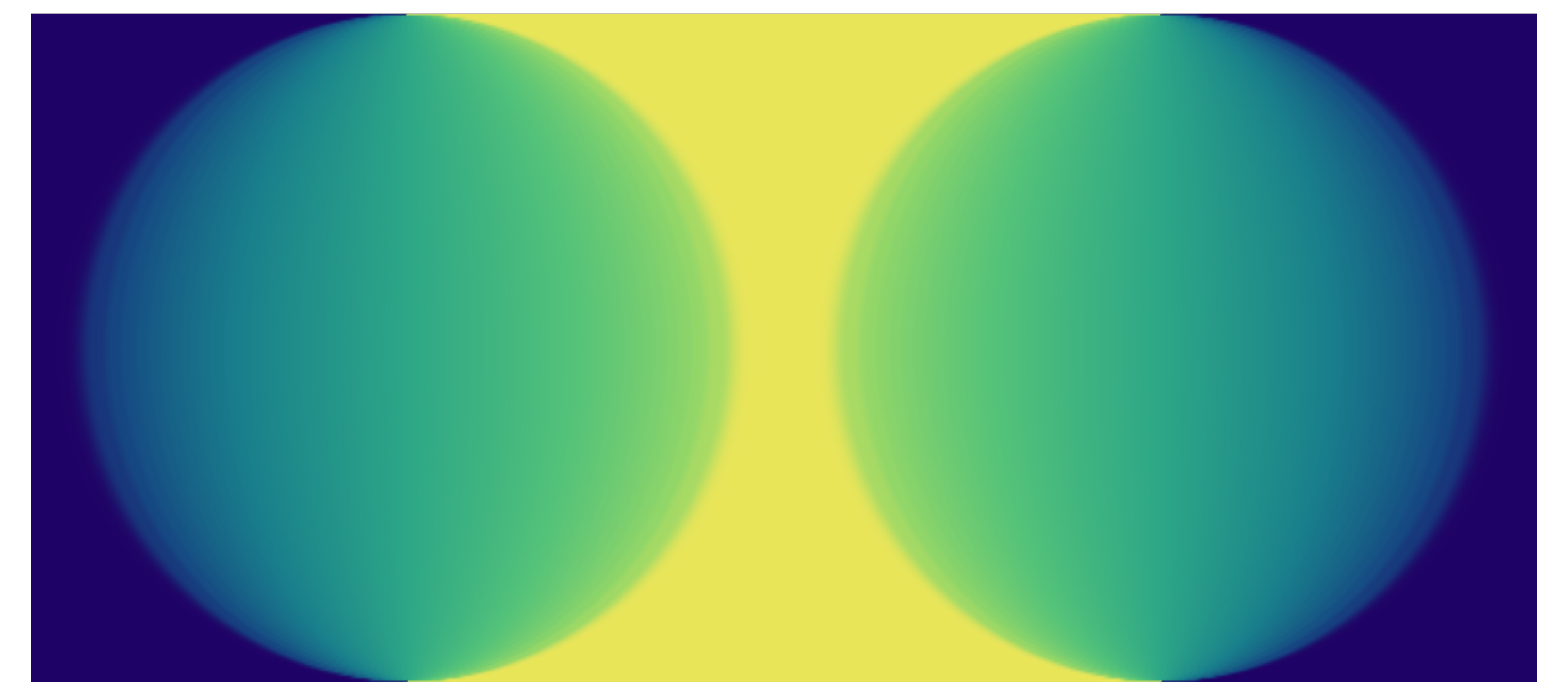}};

\draw[ultra thick] (1.73*1.125,-1.8) -- (1.73*1.125,-1.6);
\draw (1.73*1.125,-2) node {$\lambda$};
\draw[ultra thick] (-1.73*1.125,-1.8) -- (-1.73*1.125,-1.6);
\draw (-1.73*1.125,-2) node {$-\lambda$};
\draw[ultra thick] (-3.86,-1.7) -- (3.86,-1.7);
\draw[densely dashed,thick] (-4.1,0) -- (4.1,0);
\draw (0,2) node{(c)};
\draw[ultra thick] (-3.95,-1.7) -- (-3.8,-1.7);
\draw[ultra thick] (-3.95,1.7) -- (-3.8,1.7);
\draw (-4.2,1.75) node {$\phantom{+}1$};
\draw (-4.2,-1.75) node {$-1$};
\end{scope}
\end{tikzpicture}
\caption{Density profiles associated to the model with $\alpha=0$, for different values of $\lambda$. The numerical data is shown for $R=256$. (a): $\lambda=1$, (b): $\lambda=7/8$, (c): $\lambda=9/8$. Color code: blue means zero density, yellow is unit density, and intermediate colors interpolate between the two values. $Y=y/R$ is the vertical imaginary time coordinate, with boundary conditions imposed by the state $\ket{N}$ at $Y=\pm 1$. For $\lambda>1$ the two fluctuating disks are independent, while they touch for $\lambda=1$ at the point $Y=0,X=0$.}
\label{fig:limitshapes_alpha0}
\end{figure}
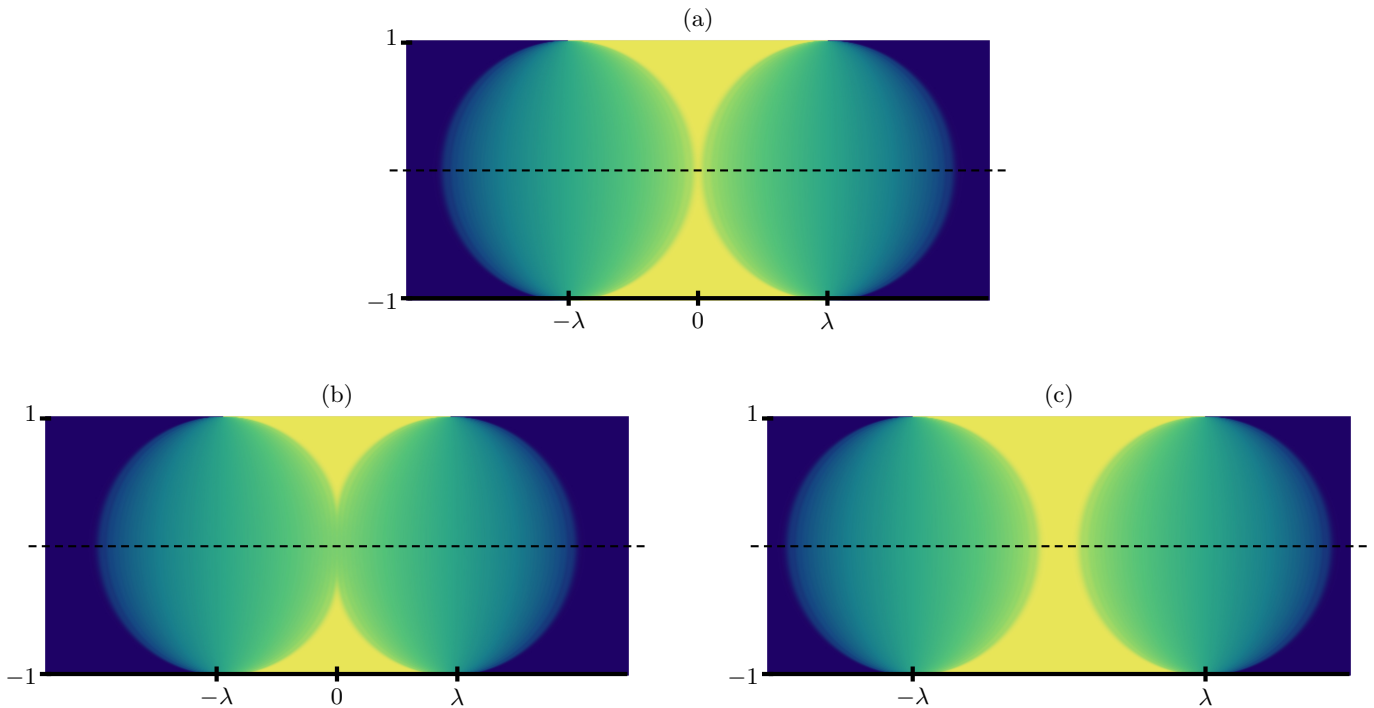

We define the partition function as the following matrix element of the imaginary time propagator of the free, one-dimensional, lattice fermions described above
\begin{equation}\label{eq:partitionfunctiondefinition}
	Z_N(R)=\bra{N}e^{2RH}\ket{N},
\end{equation}
where
\begin{equation}\label{eq:boundarystate}
	\ket{N}=\ket{2L+1}:=\prod_{x\in I_L} c_x^{\dag}\ket{0}, \qquad I_L:=\big[-L,\dots,L\big].
\end{equation}
$\ket{0}$ is the fermionic vacuum, annihilated by all $c_j$'s. 
The state $\ket{N}$ corresponds to fully occupied integer sites between the positions $-L$ and $L$, see  Fig.~\ref{fig:boundarystate}.
Then, the expectation value of any local observable $\hat{O}_x$ is defined as
\begin{align}\label{eq:observables}
    \braket{\hat{O}_x}_{y,R}:=\frac{\braket{N|e^{(R-y)H}\hat{O}_x e^{(R+y)H}|N}}{Z_N(R)},
\end{align}
where $R\geq 0$ and $y\in[-R,R]$ is a vertical coordinate. Said differently, we consider imaginary time evolution starting from $\ket{N}$ at time $-R$, and conditioned to go back to $\ket{N}$ at time $R$. For example, substituting $\hat{O}_x=c_x^\dag c_x$ in \eqref{eq:observables} yields the average density profile at position $(x,y)$, see for example Figure~\ref{fig:limitshapes_alpha0}.

For $\alpha=0$, the model is precisely the one studied in \cite{Pallister_2022}, and it can be mapped onto a system of non-intersecting random walks studied in \cite{Adler2013}. The case $\alpha>0$ can be seen as a generalization of those references. However, the partition function itself can be related to random matrix and Coulomb gas models which have been studied before that \cite{GrossWitten,Wadia,Jurkiewicz1983}, see \cite{Pallister_2022,Walsh2023} for a discussion of the mapping.

The basic limit we will be first interested in is the scaling (or hydrodynamic) limit where both particle number $N$ and width $R$ are large, that is $R\to\infty,N\to\infty$, with ratio
\begin{equation}\label{eq:hydrolimit}
    \lambda=\frac{N}{2R}
\end{equation}
fixed to some finite value (said differently fix some $\lambda>0$, set $R=N/(2\lambda)$ and send $N\to\infty$). In such a limit, and in terms of the rescaled variables $X=x/R,Y=y/R$, the model for $\alpha=0$ is known to exhibit spatial phase separation between fully ordered phases with zero (or unit) density and a fluctuating phase with non-trivial density profiles. 
These profiles are shown in Figure \ref{fig:limitshapes_alpha0} for several values of $\lambda$. For $\lambda>1$ one gets two independent fluctuating regions which are disks $(X\pm \lambda)^2+Y^2\leq 1$. Precisely for $\lambda=1$ the two regions touch, and for $\lambda<1$ they merge into a single region which can also be determined. The phenomenon was called merger transition in \cite{Pallister_2022}, and tacnode in \cite{Adler2013}.

The $\alpha$-generalized density profiles are not known and will be computed in the next section. For technical reasons--and unless specified otherwise--we will assume that $0\leq \alpha\leq 1/8$. In the next subsection we begin with a discussion of free energy which is simpler than correlations, but already contains nontrivial information.
\subsection{Free energy and phase transition}
The fact that the Hamiltonian is free-fermionic and translation invariant allows to express $Z_N(R)$ as a Toeplitz determinant. Indeed a straightforward application of Wick's theorem shows
\begin{align}\label{eq:Z_toeplitz}
    Z_N(R)=\det T_N(e^{2R\varepsilon})
\end{align}
where $T_N(e^{\tau \varepsilon})$ is the $N\times N$ matrix with elements
\begin{align}\label{eq:Tdef}
    T_N(e^{\tau \varepsilon})_{xx'}&=\braket{0|c_x e^{\tau H}c_{x'}^\dag e^{-\tau H}|0}\\\label{eq:Tdef2}
   &= \int_{-\pi}^{\pi} \frac{dk}{2\pi}e^{-\ci k(x-x')}e^{\tau \varepsilon(k)}
\end{align}
and $\eps(k)$ is the dispersion \eqref{eq:dispersion} corresponding to $H$. Notice that since the matrix elements are Toeplitz (depend only on $x-x'$) it does not really matters where the indices might start. The asymptotic behavior of the determinant can be evaluated in the hydrodynamic limit using known techniques, and in fact the result can even be extracted from Ref.~\cite{Jurkiewicz1983}.

Instead of just quoting the result we will derive it in a way which will prove useful later on. First, notice that by Szeg\H{o}'s theorem \cite{Szego} (see \eqref{eq:szegothm} and  appendix~\ref{app:gcboformula}) we have
\begin{equation}\label{eq:z_simplelimit}
    \lim_{N\to \infty} Z_N(R)=\exp\left(R^2\left[1+2\alpha^2\right]\right).
\end{equation}
Formally this is the hydrodynamic limit at $\lambda=\infty$. With this in mind, we define the bulk free energy as
\begin{equation}\label{eq:freenrjdef}
    f(\lambda,\alpha)=-\lim_{\lambda} \frac{1}{4R^2}\log\left[e^{-R^2(1+2\alpha^2)} Z_N(R)\right],
\end{equation}
where $\lim_{\lambda}$ is a shorthand for the hydrodynamic limit corresponding to a given $\lambda$, see \eqref{eq:hydrolimit}.

Next it is possible to write for fixed $R$
\begin{equation}\label{eq:Zrecursion}
    \frac{Z_{N+1}(R)Z_{N-1}(R)}{Z_N(R)^2}=1-u_N^2
\end{equation}
where the $u_N$'s satisfy the recursion relation \cite{ClioCresswell_1999,Adler2003,ChouteauTarricone}
\begin{equation}\label{eq:discretePIImaintext}
\frac{N u_N}{R(1-u_N^2)}=u_{N-1}+u_{N+1}-2\alpha \left(u_{N-2}\left[1-u_{N-1}^2\right]-u_N(u_{N-1}+u_{N+1})^2 +u_{N+2}\left[1-u_{N+1}^2\right]\right)
\end{equation}
which belongs to the discrete Painlevé II hierarchy. We use conventions where $u_0=1$ and $Z_0(R)=1$. For completeness a derivation of this relation is provided in Appendix \ref{app:asymptotics_Bessel}.
In the hydrodynamic limit we expect $u_N\to u$, $u_{N\pm 1}\to u$, $u_{N\pm 2}\to u$ where $u=u(\lambda,\alpha)$ is a solution to the fixed point equation
\begin{equation}
    \frac{\lambda u}{1-u^2}
    =u-2\alpha u(1-3u^2)
\end{equation}
which belongs to $[0,1]$. Now, Taylor expansion of the lhs of \eqref{eq:Zrecursion} to second order  yields
\begin{align}
    \partial_\lambda^2 f(\lambda,\alpha)=-\log [1-u^2(\lambda,\alpha)],
\end{align}
where \footnote{For $\alpha=0$ we simply have $u(\lambda,0)=\sqrt{1-\lambda}$ if $\lambda\leq 1$, $u(\lambda,0)=0$ otherwise.}
\begin{align}\label{eq:ulambda}
    u^2(\lambda,\alpha)=
    \begin{cases}
\frac{-1+8\alpha+\sqrt{(1+4\alpha)^2-24\alpha \lambda }}{12\alpha}\qquad &,\qquad \lambda\leq \lambda_c\\
0\qquad &,\qquad \lambda\geq \lambda_c
    \end{cases}
\end{align}
which notice is continuous at 
\begin{equation}
    \lambda_c=1-2\alpha.
\end{equation}
Using the fact that $f(\lambda,\alpha)\to 0$ for large $\lambda$, as expected from \eqref{eq:z_simplelimit}, and continuity of the first and second derivatives at $\lambda=\lambda_c$ yields an explicit but complicated expression for the free energy, which we do not write here. However it is worth noting that it is identically zero for $\lambda\geq \lambda_c$. What is most interesting is the behavior of the free energy near $\lambda_c$, from below:
\begin{align}\label{eq:thirdorder}
    f(\lambda,\alpha)&\underset{\lambda\to\lambda_c^-}{\sim}\frac{(\lambda_c-\lambda)^3}{6(1-8\alpha)} \qquad,\qquad \alpha<1/8.
\end{align}
This result can be interpreted as a third order phase transition \cite{Periwal,Jurkiewicz1983}, since the third derivative is discontinuous at $\lambda=\lambda_c$ (for $\alpha=0$ this is the known Gross-Witten-Wadia  transition \cite{GrossWitten,Wadia}). In the specific case $\alpha=1/8$ the behavior is different, and we obtain \cite{Periwal}
\begin{align}\label{eq:fivehalforder}
    f(\lambda,1/8)\underset{\lambda\to\lambda_c^-}{\sim}  \frac{8 (\lambda_c-\lambda)^{5/2}}{15\sqrt{3}}.
\end{align}
Let us go back to the case $\alpha=0$, for which $\lambda_c=1$. It is clear that the pictures shown in Figure~\ref{fig:limitshapes_alpha0} provide a physical interpretation for the third order transition. Indeed the free energy is that of the fluctuating region(s), since the trivial frozen region should not contribute. Then for $\lambda>\lambda_c=1$ the two regions become statistically independent so the free energy does not depend on $\lambda$. For $\lambda<1$ this is no longer the case. Precisely at $\lambda=1$ the two regions touch, and the subtle interaction between the two generates the phase transition. 

The main goal of this paper is to study the case $\alpha>0$, and in particular the transition for $\alpha=1/8$ which due to \eqref{eq:fivehalforder} is expected to be of a different--higher order--nature. Our ultimate objective is to study correlations close to the merger point $X=0$, $Y=0$ in this case. Let us also mention that the partition function $Z_N(R)$ has also been studied in case $R$ is pure imaginary (real time), but leads to a different behavior \cite{Krapivsky_2018,PerezGarcia2024}.

\section{Correlations and limit shapes}
\label{sec:correlations}
\subsection{Exact formulas for the two-point function}
Let us summarize various exact formulas for the two point function. These formulas follow from the free fermionic nature of the problem, combined with known results on the theory of orthogonal polynomials. They go, roughly speaking, with increasing order of complexity. Of course, all higher order correlations can be reduced to this one thanks to Wick's theorem. Before describing them let us point out that they inherit many symmetries from the underlying model, in particular a symmetry $x\leftrightarrow x'$ and a symmetry $x,x' \to -x, -x'$, and a symmetry $y\to -y$.

\begin{figure}[htbp]
\begin{tikzpicture}
\node at (0,0) {\includegraphics[width=0.45\textwidth]{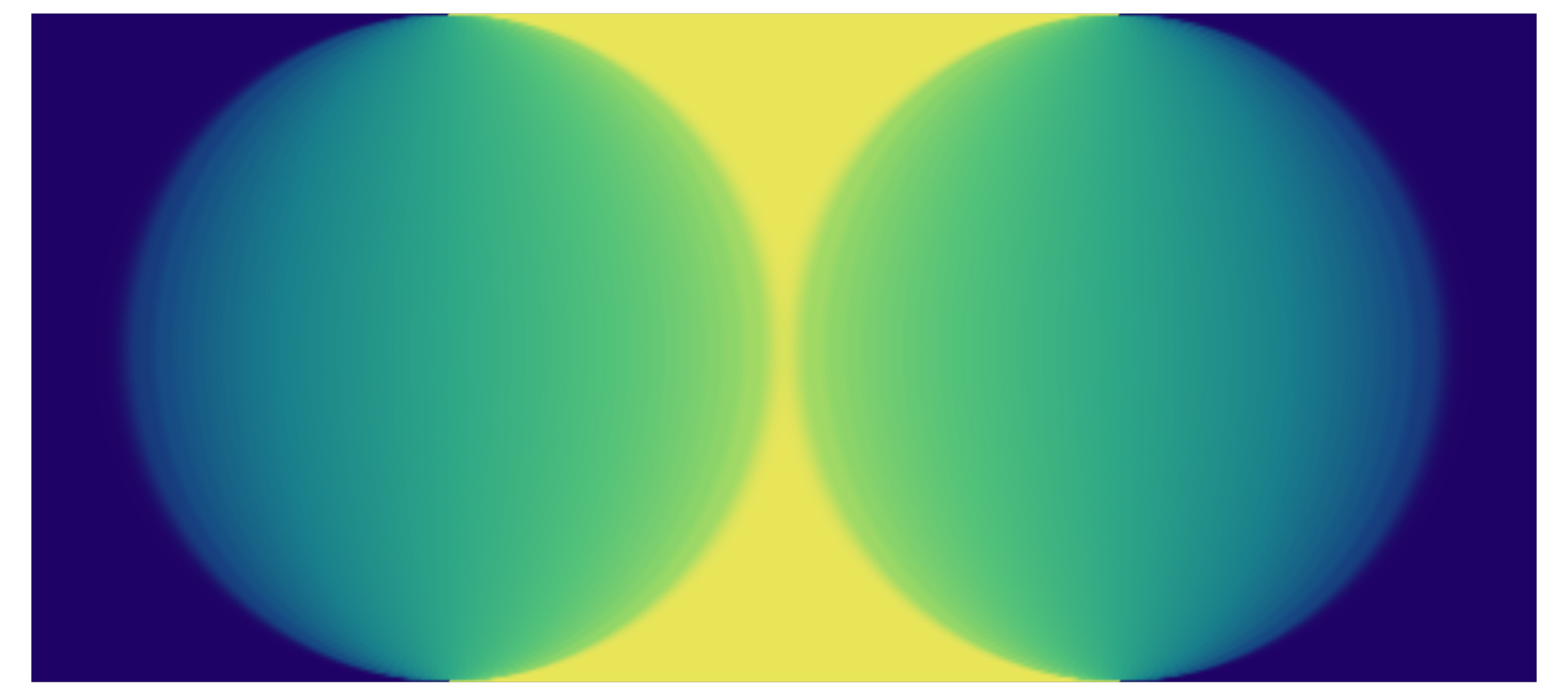}};
\draw (0,2) node {(a)};
\draw[ultra thick] (-3.95,-1.7) -- (-3.8,-1.7);
\draw[ultra thick] (-3.95,1.7) -- (-3.8,1.7);
\draw (-4.2,1.75) node {$\phantom{+}1$};
\draw (-4.2,-1.75) node {$-1$};
\draw[ultra thick] (-2*1.73,-1.8) -- (-2*1.73,-1.6);
\draw (-2*1.73,-2) node {$-2$};
\draw[ultra thick] (-1.73,-1.8) -- (-1.73,-1.6);
\draw (-1.73,-2) node {$-\lambda_c$};
\draw[ultra thick] (0,-1.8) -- (0,-1.6);
\draw (0,-2) node {$0$};
\draw[ultra thick] (1.73,-1.8) -- (1.73,-1.6);
\draw (1.73,-2) node {$\lambda_c$};
\draw[ultra thick] (2*1.73,-1.8) -- (2*1.73,-1.6);
\draw (2*1.73,-2) node {$2$};
\draw[ultra thick] (-3.86,-1.7) -- (3.86,-1.7);
\draw[densely dashed,thick] (-4.1,0) -- (4.1,0);
\begin{scope}[xshift=9.5cm]
\node at (0,0) {\includegraphics[width=0.45\textwidth]{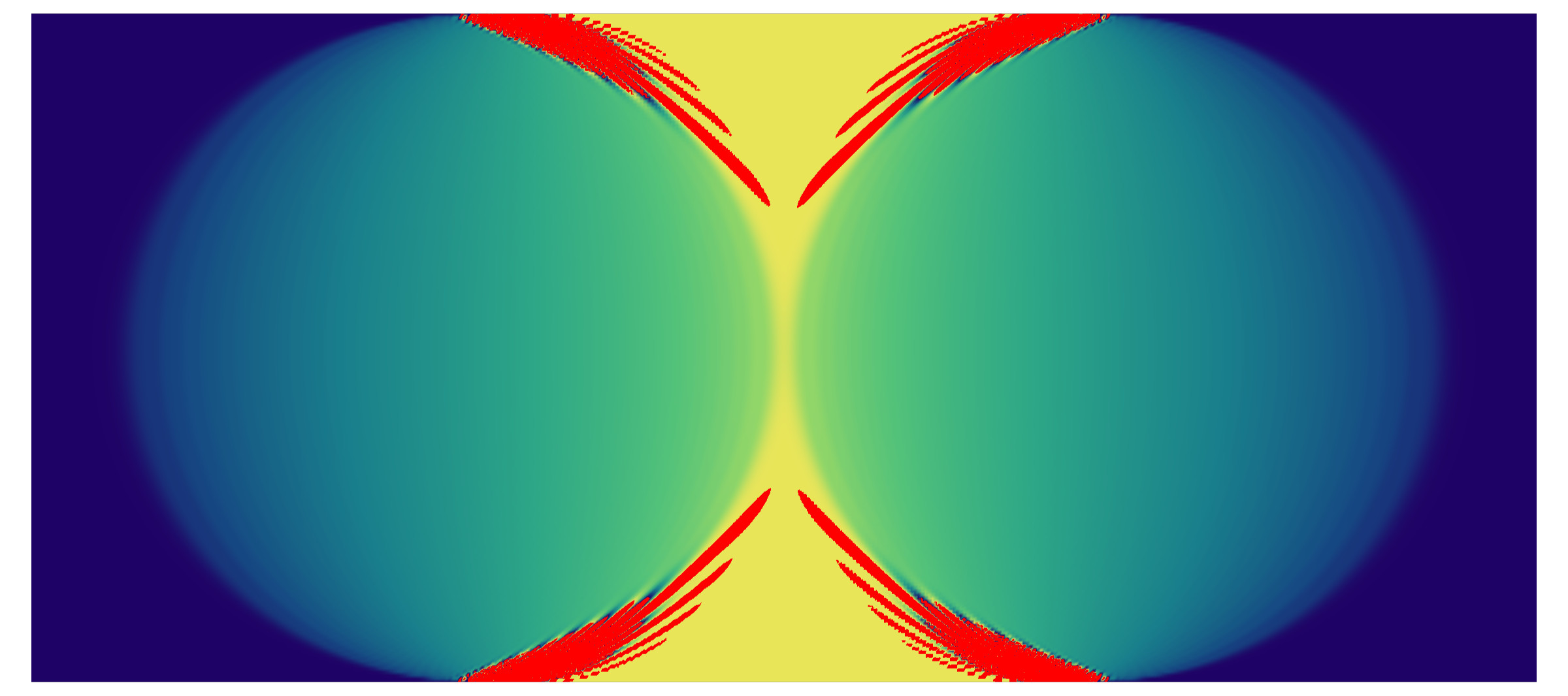}};
\draw (0,2) node {(b)};
\draw[ultra thick] (-3.95,-1.7) -- (-3.8,-1.7);
\draw[ultra thick] (-3.95,1.7) -- (-3.8,1.7);
\draw (-4.2,1.75) node {$\phantom{+}1$};
\draw (-4.2,-1.75) node {$-1$};
\draw[ultra thick] (-2*1.73,-1.8) -- (-2*1.73,-1.6);
\draw (-2*1.73,-2) node {$-2$};
\draw[ultra thick] (-0.875*1.73,-1.8) -- (-0.875*1.73,-1.6);
\draw (-0.875*1.73,-2) node {$-\lambda_c$};
\draw[ultra thick] (0,-1.8) -- (0,-1.6);
\draw (0,-2) node {$0$};
\draw[ultra thick] (1.73*0.875,-1.8) -- (1.73*0.875,-1.6);
\draw (0.875*1.73,-2) node {$\lambda_c$};
\draw[ultra thick] (-3.86,-1.7) -- (3.86,-1.7);
\draw[ultra thick] (2*1.73,-1.8) -- (2*1.73,-1.6);
\draw (2*1.73,-2) node {$2$};
\draw[densely dashed,thick] (-4.1,0) -- (4.1,0);
\end{scope}
\begin{scope}[yshift=-5cm,xshift=4.5cm]
\node at (0,0) {\includegraphics[width=0.45\textwidth]{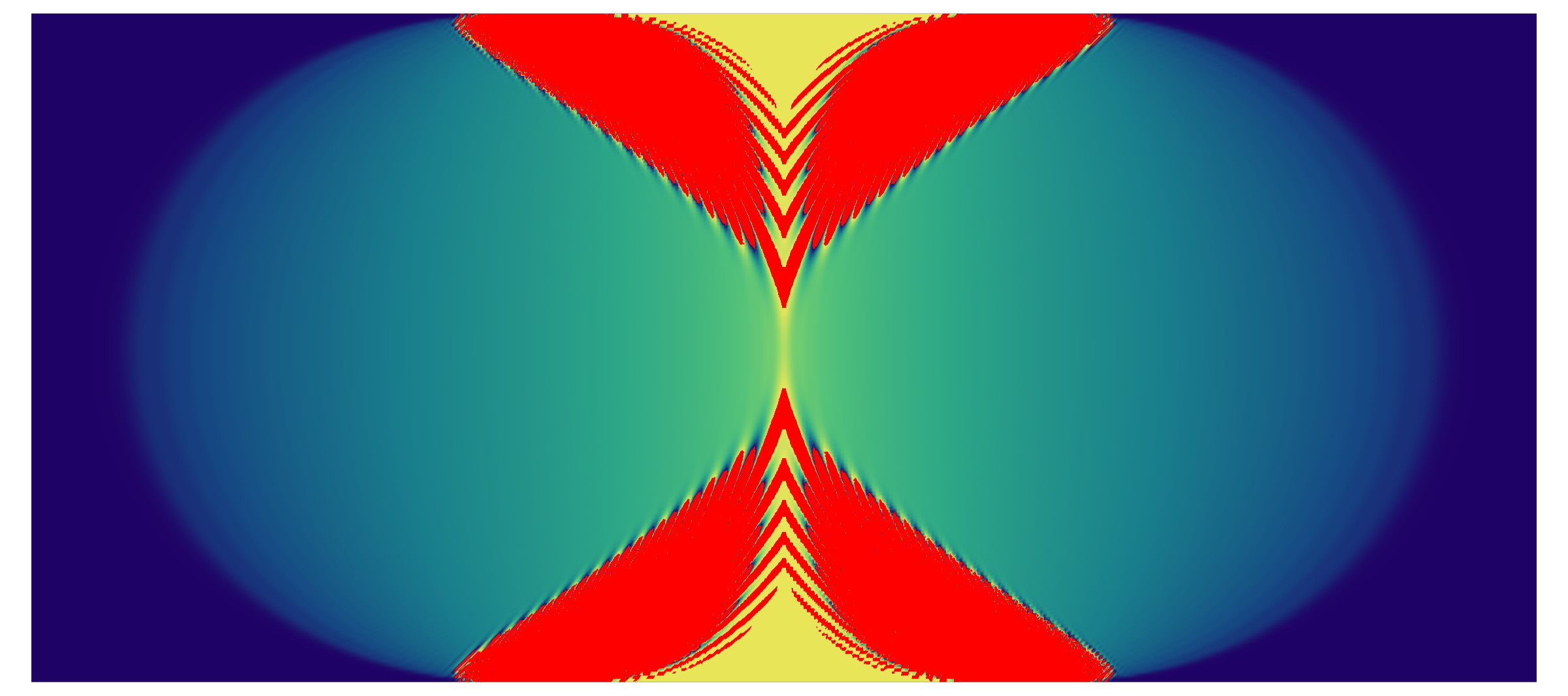}};
\draw (0,2) node {(c)};
\draw (-4.2,1.75) node {$\phantom{+}1$};
\draw[ultra thick] (-3.95,-1.7) -- (-3.8,-1.7);
\draw[ultra thick] (-3.95,1.7) -- (-3.8,1.7);
\draw (-4.2,-1.75) node {$-1$};
\draw[ultra thick] (-2*1.73,-1.8) -- (-2*1.73,-1.6);
\draw (-2*1.73,-2) node {$-2$};
\draw[ultra thick] (-0.75*1.73,-1.8) -- (-0.75*1.73,-1.6);
\draw (-0.75*1.73,-2) node {$-\lambda_c$};
\draw[ultra thick] (0,-1.8) -- (0,-1.6);
\draw (0,-2) node {$0$};
\draw[ultra thick] (1.73*0.75,-1.8) -- (1.73*0.75,-1.6);
\draw (0.75*1.73,-2) node {$\lambda_c$};
\draw[ultra thick] (-3.86,-1.7) -- (3.86,-1.7);
\draw[ultra thick] (2*1.73,-1.8) -- (2*1.73,-1.6);
\draw (2*1.73,-2) node {$2$};
\draw[densely dashed,thick] (-4.1,0) -- (4.1,0);
\end{scope}
\end{tikzpicture}
\caption{Density profile corresponding to the imaginary time evolution starting from the initial state. In all three pictures $\lambda$ is set to the critical merger value $\lambda_c=1-2\alpha$. (a) $\alpha=0$ (b) $\alpha=1/16$ (c) $\alpha=1/8$. In the (a) and (b) cases we will see that the density goes to one with square-root behavior, $\sqrt{X}$. For (c) the singularity is $X^{1/4}$, so sharper.}
\label{fig:densitiesalpha}
\end{figure}
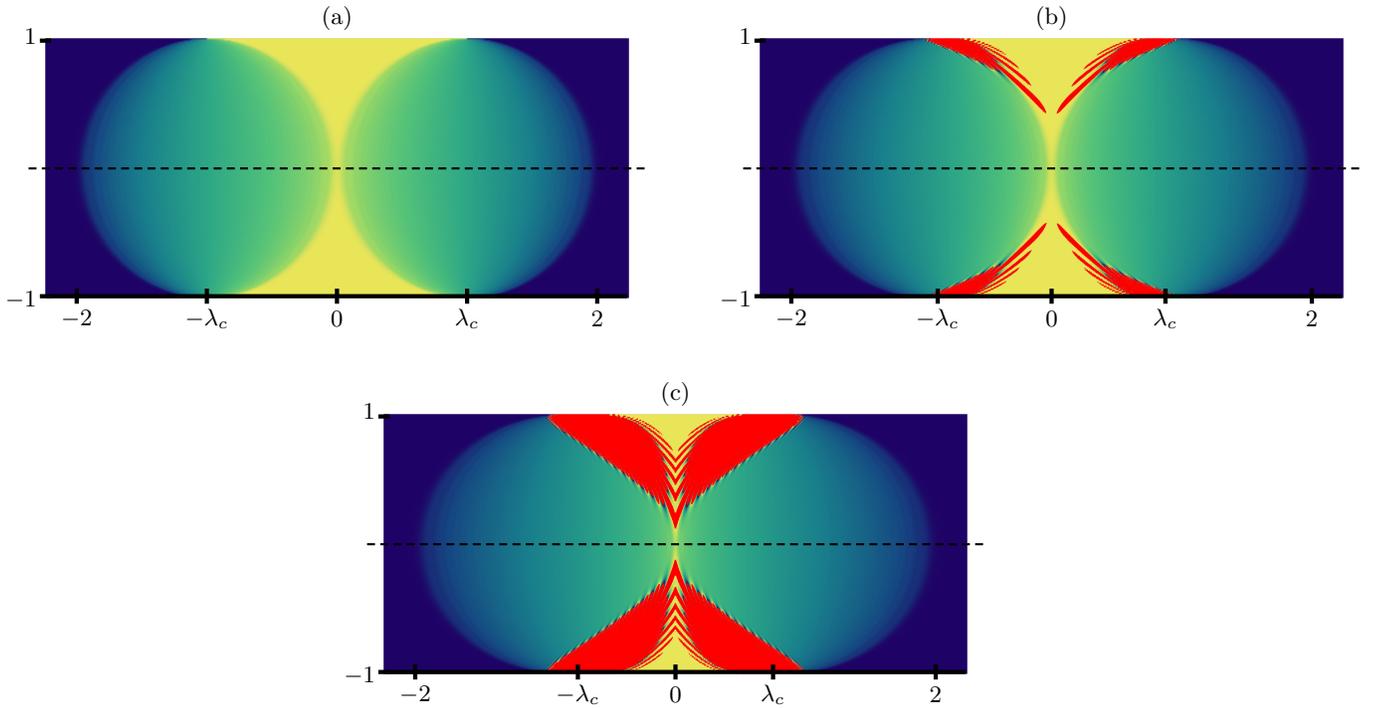

\begin{itemize}
    \item The first one is 

\begin{equation}\label{eq:exactinverse}
    \braket{c_x^\dag c_{x'}}_{y,R}=\sum_{m,n=-L}^L T_N(e^{(R+y)\varepsilon})_{x'm}T_N^{-1}(e^{2R \varepsilon})_{mn} T_N(e^{(R-y)\varepsilon})_{n,x'}
\end{equation}
and it follows from Wick's theorem, combined with the identity $\det A/\det B=\det(B^{-1}A)$ and the fact that $A$ is a simple perturbation of $B$, see Appendix~\ref{app:freefermions}. Recall $T_N(e^{\tau \varepsilon})$ is the $N\times N$ Toeplitz matrix defined in \eqref{eq:Tdef2}. This equation can be used for numerics, but while the implementation is very easy it is not the most efficient, as it requires the inversion of a rather large ill-conditioned matrix, which requires huge numerical precision. Nevertheless, we used it to generate the pictures shown in Fig.~\ref{fig:densitiesalpha}. We will be most interested in correlations at imaginary time $y=0$ (dashed line in the figure), so we restrict for convenience to this case in the following.
\item The second one makes use of the theory of orthogonal polynomials. To understand how those come about, write the time evolved state (recall $N=2L+1$)
\begin{align}
    \ket{\psi_N(R)}=e^{RH}\ket{N}=c_{-L}^\dag (R)\ldots c_L^\dag(R)\ket{0}
\end{align}
in terms of the Heisenberg picture $c_x^\dag(\tau)=e^{\tau H}c_x^\dag e^{-H\tau}$. In Fourier space those read
\begin{align}
    c_x^\dag(R)=\int_{-\pi}^{\pi} \frac{dk}{2\pi} e^{-\ci k x}e^{R\varepsilon(k)}d^\dag(k).
\end{align}
Let us introduce a modified set of operators $f_l^\dag$, where each $f_l^\dag$ is some linear combination of $c_{-L}^\dag(R)$, $c_{-L+1}^\dag(R)$, up to $c_{-L+l}^\dag(R)$, for $l\in \{0,1,\ldots,N-1\}$. It is easy to check that the states 
\begin{equation}\label{eq:samestate}
\ket{\chi_N(R)}=f_0^\dag \ldots f_{N-1}^\dag \ket{0}    
\end{equation}
and $\ket{\psi_N(R)}$ are proportional for any choice of linear combination. In Fourier space
\begin{align}
    f_l^\dag=\int_{-\pi}^{\pi} \frac{dk}{2\pi} e^{\ci k L}P_l(e^{-\ci k })e^{R\varepsilon(k)}d^\dag(k)
\end{align}
where $P_l(z)$ is a polynomial of degree $l$ in $z$. Now, a clever choice is to ask that the $f_l$ satisfy anticommutation relations, and this leads to
 \begin{align}
  \delta_{lm}&=\{f_l,f_m^\dag\}   \\ \label{eq:whyortho}
  &=\int_{-\pi}^{\pi} \frac{dk}{2\pi} P_l(e^{\ci k})P_m(e^{-\ci k}) e^{2R\varepsilon(k)},
 \end{align}
so those relations are guaranteed if the $P_l(z)$ form a set of orthonormal polynomials on the unit circle (see appendix \ref{app:orthopoly} for an introduction), and we make this choice in the following. Back in real space
\begin{align}
    f_l^\dag=\sum_{x\in\mathbb{Z}} \phi_l(x)c_x^\dag,
\end{align}
 where the
 \begin{align}\label{eq:phik}
     \phi_l(x)=\int \frac{dk}{2\pi} e^{\ci k(L+x)}P_l(e^{-\ci k})e^{R\varepsilon(k)}
 \end{align}
 can be seen as single-particle wave functions. Hence
 \begin{align}
    \braket{c_x^\dag c_{x'}}_{0,R}=\braket{\chi_N(R)|c_x^\dag c_{x'}|\chi_N(R)} 
 \end{align}
 and  anticommuting the $f^\dag$ to the left yields
\begin{equation}\label{eq:singleparticles}
    \braket{c_x^\dag c_{x'}}_{0,R}=\sum_{l=0}^{N-1} \phi_l(x)\phi_l(x'),
\end{equation}
which is our second main formula. An alternative derivation relies on an exact formula for the inverse $T^{-1}(2R)$ which can be found in appendix \ref{app:freefermions}, and it also relies on orthogonal polynomials. See also \cite{Adler2013} for $\alpha=0$.

Formula \eqref{eq:singleparticles} is useful for numerics, because the $\phi_l(x)$ satisfy two recursion relations which stem from known properties of the polynomials. The first one follows from the famous Szeg\H{o} recursion \eqref{eq:Szegorecursion}, and yields
\begin{equation}
   \sqrt{1-u_{l+1}^2} \phi_{l+1}(x)=\phi_l(x-1)+(-1)^{l+1} u_{l+1} \phi_l(l-2L-x).
\end{equation}
This allows to build recursively all $\phi_{k}$ from the numerical knowledge of $\phi_0(x)\propto \int \frac{dk}{2\pi}e^{-\ci k(x+L)}e^{R\varepsilon(k)}$.
The second one follows from integrating by parts the contour integral using the formulas \eqref{eq:dpdz},\eqref{eq:bess_an},\eqref{eq:bess_bn} for $\frac{dP_l(z)}{dz}$, see appendices \ref{app:orthopoly} and \ref{app:asymptotics_Bessel}. The final result is complicated for $\alpha> 0$, so for simplicity we only provide the case $\alpha=0$,
\begin{equation}\label{eq:recursion_phi}
    \frac{L+x}{R}\phi_l (x)=\frac{\phi_l (x-1)+\phi_l (x+1)}{2}+\frac{l+R u_{l}u_{l+1}}{R}\phi_l (x)-(-1)^l\left[u_{l}\phi_l (l-2L-x-1)+u_{l+1}\phi_l (l-2L-x)\right].
\end{equation}
This shows that the single-particle wave functions are not directly that of a free quantum mechanical diagonalization problem, in the sense that they do not come from a simple eigenvalue problem. This is because of the last two terms on the rhs. However in situations where the $u_l$ can be neglected this will be again the case, and we recover discussions which can be found in  \cite{Eisler_2009,Walsh2023}. 
\item The third one takes the form of a double contour integral formula
\begin{align}\label{eq:generaldoublecontourintegral}
    \braket{c_x^\dag c_{x'}}_{0,R}&=\int_{-\pi}^{\pi}\frac{dk}{2\pi}\int_{-\pi}^{\pi}\frac{dq}{2\pi} e^{\ci (k x-q x')+R(\varepsilon(k)+\varepsilon(q))} \frac{e^{\ci (L+1)(q-k)}P_N(e^{\ci k}) P_N(e^{-\ci q})-e^{-\ci L (q-k)}P_N(e^{-\ci k})P_N(e^{\ci q})}{1-e^{\ci (q-k)}},
\end{align}
and follows from \eqref{eq:singleparticles},\eqref{eq:phik} combined with the Christoffel-Darboux formula \eqref{eq:cd_def},\eqref{eq:cd_formulabis}. As we shall see in the next section, it is particularly useful to study the hydrodynamic limit. 
\item The last one is specifically designed to study the center edge limit, which will be investigated in section \ref{sec:edge}. First introduce the semi-infinite matrix $K_N$, with elements
\begin{align}\label{eq:basickern}
    (K_N)_{jl}
    &=\sum_{p\geq 0} J_{N+j+p+1}^{(\alpha)}(2R) J_{N+l+p+1}^{(\alpha)}(2R)
\end{align}
for $j,l\geq 0$, where

\begin{align}
   J_n^{(\alpha)}(R)= \int_{-\pi}^{\pi} \frac{dk}{2\pi}e^{-\ci k n}e^{\ci R [\sin k-\alpha \sin 2k]}
\end{align}
is a deformation of the usual Bessel function (notice it is real). 
Then we have the many Bessel formula
\begin{align}\label{eq:rankoneperturbation}
    \braket{c_x^\dag c_{x'}}_{0,R}&=
    \sum_{p\geq 0} J^{(-\alpha)}_{x+p-L}(R) J^{(-\alpha)}_{x'+p-L}(R)-(-1)^{x-x'}\sum_{j,l\geq 0}(1-K_N)^{-1}_{jl} a_x(j)a_{x'}(l),
\end{align}
where
\begin{align}\label{eq:axjdef}
     a_x(j)&=J_{j+L+x+1}^{(\alpha)}(R)-\sum_{p\geq 0} J^{(\alpha)}_{j+N+p+1}(2R)
    J_{L+p+1-x}^{(\alpha)}(R),
\end{align}
and $(1-K_N)^{-1}$ is the inverse of the semi-infinite matrix $1-K_N$. 

It is worth noting that the first term in \eqref{eq:rankoneperturbation} coincides exactly with the two point function in the domain wall geometry found in \cite{Bocini2021}. The second term can then be interpreted as a rank one perturbation of this kernel, which may or may not be relevant in a given scaling limit.
\end{itemize}
\textbf{Derivation of equation \eqref{eq:rankoneperturbation}.}
The technique to find it is explained in \cite{Adler2014,Johansson_review}, we recall it here for completeness. The starting point is an expression for the Christoffel-Darboux kernel in terms of Toeplitz determinants
\begin{align}
    \sum_{m=0}^{N-1} P_m(z)P_m(1/w)&=\left(\frac{z}{w}\right)^{N-1} \frac{D_{N-1}[(1-e^{\ci k}/z)(1-w e^{-\ci k})e^{2R\varepsilon(k)}]}{D_N[e^{2R \varepsilon(k)}]}
\end{align}
where recall $D_n[h(k)]=\det_{1\leq i,j\leq n} \int \left(\frac{dk}{2\pi}e^{-\ci k(i-j)}h(k)\right)$ is the $n\times n$ Toeplitz determinant corresponding to the symbol $h(k)$. 
We will now make use of the Geronimo-Case \cite{GeronimoCase}, Borodin-Okounkov \cite{BorodinOkounkov} formula, see Appendix~\ref{app:gcboformula} and \eqref{eq:gc}, to express both Toeplitz determinants as Fredholm determinants:
\begin{align}\label{eq:ratiodettofred}
   \frac{D_{N-1}[(1-e^{\ci k}/z)(1-w e^{-\ci k})e^{2R\varepsilon(k)}]}{D_N[e^{2R \varepsilon(k)}]}&=\frac{e^{-R(1/z+\alpha/z^2+w+\alpha w^2)}}{1-w/z}\frac{\det(1-K_{N-1}[1/z,w])}{\det(1-K_N[0,0])}
\end{align}
with kernel acting on sequences in $\ell^2(\mathbb{N})$ given by
\begin{align}
    K_n[1/z,w]_{jl}=\oint' \frac{d\zeta}{2\ci\pi \zeta^{j+n+1}}\oint' \frac{d\omega}{2\ci\pi\omega^{-l-n}}\frac{e^{R(g_*(\zeta)-g_*(\omega))}}{\zeta-\omega}
    \frac{1+w/\zeta}{1+\zeta/z}\frac{1+\omega/z}{1+w/\omega}
\end{align}
with
\begin{align}
    g_*(\zeta)=\zeta-\frac{1}{\zeta}-\alpha\left(\zeta^2-\frac{1}{\zeta^2}\right)
\end{align}
and counterclockwise contours $\oint' d\zeta \oint' d\omega$ such that $|w|<|\omega|<|\zeta|<|z|$. 
Now write
\begin{align}
      \frac{1+w/\zeta}{1+\zeta/z}\frac{1+\omega/z}{1+w/\omega}=\frac{\omega}{\zeta}\left(1-\frac{(w-z)(\zeta-\omega)}{(z+\zeta)(w+\omega)}\right),
\end{align}
which implies the marvelous simplification
\begin{align}
    K_n[1/z,w]_{jl}=K_{n+1}[0,0]_{jl}-(w-z) \left(\oint_{|\zeta|<|z|} \frac{d\zeta e^{R g_*(\zeta)}}{2\ci\pi \zeta^{j+n+2}(z+\zeta)}\right)\left(\oint_{|w|<|\omega|} \frac{d\omega e^{-R g_*(\omega)}}{2\ci\pi \omega^{-l-n-1}(w+\omega)}, \right)
\end{align}
which means the kernel appearing in the numerator in \eqref{eq:ratiodettofred} is a rank one perturbation of $K_N[0,0]$. Hence
\begin{align}
    \frac{\det(1-K_{N-1}[1/z,w])}{\det(1-K_N[0,0])}&=\det((1-K_N)^{-1}(1-K_{N-1}[1/z,w]))\\
    &=1-(z-w)\sum_{j,l\geq 0} (1-K_N)^{-1}_{jl} \left(\oint_{|\zeta|<|z|} \frac{d\zeta e^{R g_*(\zeta)}}{2\ci\pi \zeta^{j+N+1}(z+\zeta)}\right)\left(\oint_{|w|<|\omega|} \frac{d\omega \omega^{l+N} e^{-R g_*(\omega)}}{2\ci\pi (w+\omega)} \right),
\end{align}
where we have used $\det_{ij}(\delta_{ij}+u_i v_j)=1+\sum_{k} u_k v_k$, 
and the ratio of determinants can be written as
\begin{align}
      \frac{D_{N-1}[(1-e^{\ci k}/z)(1-w e^{-\ci k})e^{2R\varepsilon(k)}]}{D_N[e^{2R \varepsilon(k)}]}&=e^{-R\left(\frac{1}{z}+\frac{\alpha}{z^2}+w+\alpha w^2\right)}\left[\frac{1}{1-w/z}-\sum_{j,l\geq 0} (1-K_N)^{-1}_{jl} a_j[z]a_l[1/w]\right],
\end{align}
where
\begin{align}
    a_j[z]&=\oint_{|\zeta|<|z|} \frac{d\zeta e^{R g_*(\zeta)}z}{2\ci\pi \zeta^{j+N+1}(z+\zeta)}\\
    &=\frac{e^{R g_*(-z)}}{(-z)^{j+N}}+\oint_{|\zeta|>|z|}\frac{d\zeta e^{R g_*(\zeta)}z}{2\ci\pi \zeta^{j+N+1}(z+\zeta)}\\
    &=\frac{e^{R g_*(-z)}}{(-z)^{j+N}}+z\sum_{p\geq 0}(-z)^p J^{(\alpha)}_{j+N+p+1}(R).
\end{align}
We have used the residue theorem to get the second line and written $\frac{1}{z+\zeta}=\zeta^{-1}\sum_{p\geq 0} (-z/\zeta)^p$. 
Plugging this representation of the Christoffel-Darboux kernel in the correlator gives
\begin{align}
    \braket{c_x^\dag c_{x'}}_R&=\oint \frac{dz}{2\ci\pi z^{x+L+1}}\oint \frac{dw}{2\ci\pi w^{-L-x'+1}}\sum_{k=0}^{N-1}P_k(z)P_k(w^{-1})e^{\frac{R}{2}(g(z)+g(w))}\\
    &=\oint \frac{dz e^{\frac{R}{2}g(z)}}{2\ci\pi z^{x-L+1}}\oint_{|w|<|z|} \frac{dw e^{\frac{R}{2}g(w)}}{2\ci\pi w^{L-x'+1}} e^{-R\left(\frac{1}{z}+\frac{\alpha}{z^2}+w+\alpha w^2\right)}\left[\frac{1}{1-w/z}-\sum_{j,l\geq 0} (1-K_N)^{-1}_{jl} a_j[z]a_l[1/w]\right]\\
      &=\sum_{p\geq 0} J^{(-\alpha)}_{x+p-L}(R) J^{(-\alpha)}_{x'+p-L}(R)-\sum_{j,l\geq 0} (1-K_N)^{-1}_{jl}a_j(x)a_l(x')
\end{align}
where $g(z)=z+1/z+\alpha (z^2+1/z^2)$, 
\begin{align}
    a_x(j)&= \oint \frac{dz e^{\frac{R}{2}[(z-\frac{1}{z})+\alpha(z^2-\frac{1}{z^2})]}}{2\ci\pi z^{x-L+1}} a_j[z]\\
     &=(-1)^{L+x} J_{j+L+x+1}^{(\alpha)}(R)-(-1)^{L+x}\sum_{p\geq 0} J^{(\alpha)}_{j+N+p+1}(2R)
    J_{L+p+1-x}^{(\alpha)}(R),
\end{align}
and we have used $J_{-n}^{(\alpha)}(R)=J_n^{(\alpha)}(-R)=(-1)^n J_n^{(-\alpha)}(R)$. Finally relabeling $a_x(j)\to (-1)^{L+x} a_x(j)$ gives \eqref{eq:rankoneperturbation}, \eqref{eq:axjdef}.

\subsection{Limit shapes}
We aim to analyze the density profile of the model, in the hydrodynamic limit and for $\alpha>0$ (see \cite{Pallister_2022} for $\alpha=0$). We will in particular obtain a simple exact formula in the middle at $Y=0$, corresponding to the dashed line in Figure~\ref{fig:densitiesalpha} (recall $X=x/R$, $Y=y/R$ in terms of the original horizontal and vertical coordinates in \eqref{eq:observables}). This will serve as a motivation to study edge limits and a new type of universal behavior.

Before doing that it is important to realize that the inclusion of a next-nearest neighbor hopping term is not innocent at all. As is discussed in \cite{Bocini2021} in the context of the semi-infinite domain wall state, the underlying statistical mechanical model does not have positive Boltzmann weights anymore \footnote{This is because a minus sign is picked whenever a fermion at site $j$ jumps to the right of another at site $j+1$, which is permitted once the next-nearest neighbor term in $H$ is non zero}. This means the density does not necessarily belong to $[0,1]$ anymore, and can take arbitrarily large positive or negative values. In the hydrodynamic limit, one gets three types of regions: (i) frozen ones where density is either one or zero, (ii) regular fluctuating ones with density in $[0,1]$, (iii) `crazy' regions in which density does not converge. These regions are illustrated in Figure \ref{fig:densitiesalpha}. The problem of crazy regions disappears for $y=0$, because expectation values read $\braket{O_x}_R=\frac{\braket{\psi(R)| O_x |\psi(R)}}{\braket{\psi(R)|\psi(R)}}$ with $\ket{\psi(R)}=e^{R H}\ket{\psi}$, we refer to \cite{Bocini2021} for a more thorough description.

There are several ways to access the density profile. One is to generalize the purely hydrodynamic approach of \cite{Pallister_2022} to our situation. This method suffers a priori from conceptual issues when $\alpha\neq 0$, as it involves optimizing over densities which are not positive, with no a priori guarantee that it is physical. However, as we shall see, it does give the correct result when $y=0$. To demonstrate that, we will use the orthogonal polynomial framework based on the exact formula \eqref{eq:generaldoublecontourintegral}, and also perform numerical checks. Another motivation comes from the fact that it will be necessary to use orthogonal polynomials to study the edge limit in any case.

Let us sketch the main steps in the derivation, which boils down to finding saddle points. For $x'=x$ the integrant in \eqref{eq:generaldoublecontourintegral} is of the form $\frac{e^{R F(k)}e^{R G(q)}-e^{R \tilde{F}(k)}e^{R \tilde{G}(q)}}{1-e^{\ci (q-k)}}$, so the numerator is a sum of two terms which factorize as functions of $k$ and $q$, and we expect the neighborhood of the saddle point(s) to dominate the integral for large $R$. The saddle point equation for $F$ reads $\frac{dF}{dk}=0$ which reads
\begin{align}
    X-\lambda +\frac{1}{R}\frac{d\log Q_N(e^{\ci k})}{dk}=0.
\end{align}
The asymptotic behavior of the orthogonal functions $Q_N(e^{\ci k})=P_N(e^{\ci k})e^{R\varepsilon(k)}$ is studied in the Appendix \ref{app:asymptotics_Bessel}, in the hydrodynamic regime where $N=2L+1$, $L=\lambda R$. The final result depends on $\lambda$, and reads
\begin{align}
    \lim_{R\to \infty} \frac{1}{R}\frac{d\log Q_{2\lambda R}(e^{\ci k})}{dk}&=\begin{cases}
    2\lambda+\cos k+2\alpha \cos 2k&,\quad \lambda \geq \lambda_c\\\lambda+(1+2\alpha [2\cos k+\cos k_c-1])\sqrt{(1+\cos k)(\cos k-\cos k_c)} &,\quad\lambda \leq \lambda_c
    \end{cases}
\end{align}
where $k_c$ is the unique solution in $[0,\pi]$ to the equation
\begin{equation}
    \cos k_c=2u(\lambda)^2-1,
\end{equation}
$u(\lambda)$ is given by \eqref{eq:ulambda}, and $k\in[-k_c,k_c]$ is implied in the second line. The answer is invariant under $k\to -k$, which means the saddle point equation corresponding to $G(q)$ gives the same saddle point. The equation may be rewritten in the form
\begin{align}\label{eq:upsilon1}
    -X=\Upsilon(k)
\end{align}
where 
\begin{equation}\label{eq:upsilondef}
    \Upsilon(k)=\begin{cases}
    \lambda+\cos k+2\alpha \cos 2k&, \quad\lambda \geq \lambda_c\\(1+2\alpha [2\cos k+\cos k_c-1])\sqrt{(1+\cos k)(\cos k-\cos k_c)} &,\quad\lambda \leq \lambda_c
    \end{cases}.
\end{equation}
One can check that \eqref{eq:upsilon1} has a unique solution $k_F\in [0,\pi]$ provided $\alpha\leq 1/8$ and $X\leq \Upsilon(0)$, which can be written as $k_F=\Upsilon^{-1}(-X)$. A similar analysis of the second term involving $\tilde{F},\tilde{G}$ yields the equation
\begin{equation}\label{eq:upsilon2}
    X=\Upsilon(k)
\end{equation}
instead. With the saddle points at hand, a full analysis of the double integral in the hydrodynamic limit can be performed. There are subtleties related to the fact that the saddle points in $k,q$ do coincide, combined with the pole at $k=q$ in \eqref{eq:generaldoublecontourintegral}. The appropriate contour deformation necessary to perform the analysis are non trivial but by now well understood (e.g. \cite{OkounkovReshetikhin,BorodinGorinlectures,Allegra_2016,Bocini2021,Walsh2023}). The final result turns out to be remarkably simple. For $X<0$, only the first term in \eqref{eq:generaldoublecontourintegral} gives a nonzero contribution to the density profile, which is $\frac{k_F}{\pi}=\frac{1}{\pi}\Upsilon^{-1}(-X)$. For $X> 0$, only the second term is nonzero, with contribution $\frac{1}{\pi}\Upsilon^{-1}(X)$ to the density profile. This is consistent with the symmetry $X\to-X$ which is obvious in the underlying model studied. Physically, this simply means that in all cases $k_F=k_F(X)=\Upsilon^{-1}(|X|)$ can be interpreted as a position dependent Fermi momentum, hence the notation.

Putting everything together, our main conclusion is that the density profile in the hydrodynamic limit is given by
\begin{equation}\label{eq:exactdensityprofiles}
    \rho(X,Y=0)=\frac{\Upsilon^{-1}(|X|)}{\pi}
\end{equation}
provided $|X|\leq \Upsilon(0)$ if $\lambda\leq \lambda_c$, $\Upsilon(\pi)\leq |X|\leq \Upsilon(0)$ if $\lambda\geq \lambda_c$. Far to the left (right), $|X|\geq \Upsilon(0)$ and the density vanishes. Above the critical value $\lambda>\lambda_c$ the density goes to one if $|X|\leq \Upsilon(\pi)=\lambda-\lambda_c$. Notice also that the density in the center is given by $\rho(0)=k_c/\pi$, while it vanishes at $X=\pm \Upsilon(0)$.

\begin{figure}[htbp]
\includegraphics[width=0.49\textwidth]{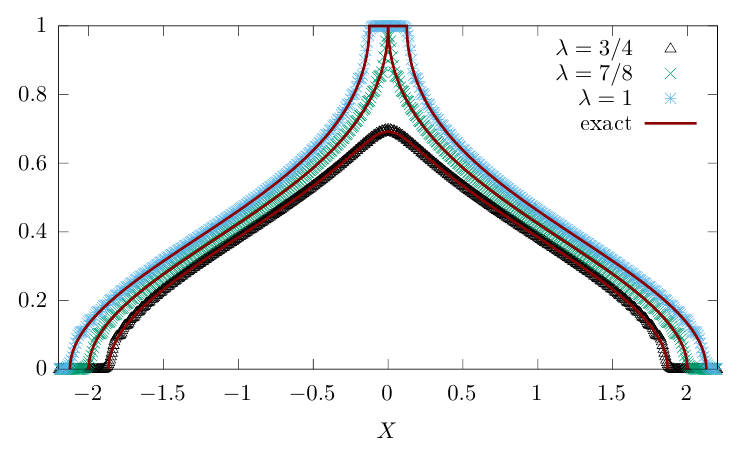}
\includegraphics[width=0.49\textwidth]{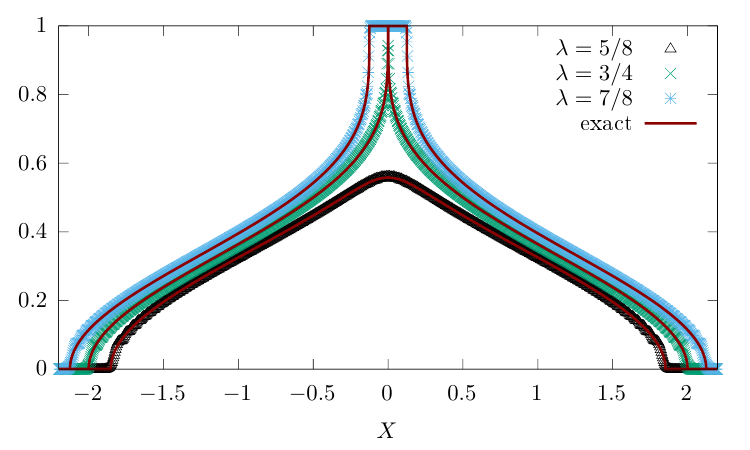}
\caption{Exact density profiles following from \eqref{eq:exactdensityprofiles}, \eqref{eq:upsilondef} as a function of $X$. Left: the numerical data for $R=128$ $\alpha=1/16$ and $\lambda=5/8,3/4,7/8$ matches perfectly the analytical result. At the critical value $\lambda_c=7/8$, the two fluctuating regions merge and (one minus) the density has a square root singularity at $X=0$. Right: same with $\alpha=1/8$.  At the critical value $\lambda_c=3/4$, the two regions merge and (one minus) the density has a fourth root singularity at $X=0$.}
\label{fig:densityyzero_plots}
\end{figure}

To better visualize all cases, the exact hydrodynamic density profiles are shown in Figure \ref{fig:densityyzero_plots} for several values of $\alpha,\lambda$.
At the merger point $\lambda=\lambda_c$, the density approaches one as
\begin{eqnarray}
    1-\rho(X)\sim \frac{1}{\pi}\sqrt{\frac{2X}{1-8\alpha}}
\end{eqnarray}
for $\alpha<1/8$ and
\begin{equation}\label{eq:fourthroot}
    1-\rho(X)\sim \frac{(8 X)^{1/4}}{\pi}
\end{equation}
for $\alpha=1/8$. This change of behavior mimics what was described in Section \ref{sec:introknown} for the free energy, see equations \eqref{eq:thirdorder} and \eqref{eq:fivehalforder}. In Section \ref{sec:edge}, we will analyze more precisely the behavior of the two point function near the merger point, and witness the appearance of a new process when $\alpha=1/8$.

\subsection{Relation to hydrodynamics}
It's important to make a comment on the relation to hydrodynamics and known particular cases.
Above the critical value $\lambda_c$ the left and right interfaces initially at $X=\pm \lambda$ are never connected through imaginary time evolution, and one gets two independent fluctuating regions. The density for $X\geq 0$ is expected to match that of an initial semi-infinite domain wall state $\ket{\psi}=\prod_{x\leq L} c_x^\dag \ket{0}$, and this can be shown directly from the alternative formula (\ref{eq:rankoneperturbation}). The corresponding density profile has been computed in \cite{Bocini2021} and indeed the two results coincide. The case $X\leq 0$ follows by symmetry. Below $\lambda\leq \lambda_c$ our result generalizes that of \cite{Pallister_2022} to $\alpha\in (0,1/8]$.

The method we used was based on orthogonal polynomials and saddle point analysis, but this is not the only way to obtain the result if one looks only for local operators such as densities or current profile. An arguably simpler method was put forward in \cite{Pallister_2022}, based on hydrodynamics \cite{Abanov_hydro}. It postulates that the Fermi momentum can be continued to the complex plane, and its imaginary time evolution is free \footnote{These equations can be thought of as the only consistent analytic continuation of the usual real time free evolution $\partial_t k+\partial_X \varepsilon(k)=0$.}, which implies $\ci\partial_Y k+\partial_X \varepsilon(k)=0$ and $-\ci\partial_Y \bar{k}+\partial_X \varepsilon(\bar{k})=0$. In such a description the real part of $k$ is ($\pi$ times) the density, while the imaginary part may be interpreted as a current. The general solution to the hydrodynamic equation can be parametrized as
\begin{align}
    X+\ci Y\varepsilon'(k)=G(k),
\end{align}
where $G$ is an analytic function. $G$ can (in principle uniquely) be determined from the boundary conditions at $Y=\pm 1$, and this is in general a nontrivial inverse problem. In our case the exact knowledge of the density profile at $Y=0$ is sufficient to find this function, and we find $G=\Upsilon$. When specifying $\alpha=0$, this equality is consistent with the density profile found in \cite{Pallister_2022}. We emphasize that the validity of hydrodynamics is not obvious when $\alpha>0$ due to the non positivity of the underlying model. This is because the hydrodynamic equations can be seen as the result of a minimization over all possible density profiles  \cite{Abanov_hydro,Stephan_lectures}, and this procedure becomes uncontrolled if unbounded non positive densities are allowed. Nevertheless it provides the correct result at least for $Y=0$, but it breaks down in the crazy regions shown in Figure \ref{fig:densitiesalpha}. 

Finally, note that our results can be generalized to the case $\alpha>1/8$ without too many extra technical difficulties. However, this is not necessary to access the various edge behaviors we are after, so we refrain from doing so.

\section{Edge behaviors}
\label{sec:edge}
Now we study more precisely the various edge behaviors, by appropriately zooming close to the place where the hydrodynamic density goes to either zero (exterior edges), or, perhaps most interestingly, one (center). 
Before doing that, we go back to the partition function.

\subsection{Scaling of the partition function near the transition}
\label{sec:freenrjneartransition}
Due to the peculiar behavior of the free energy in the hydrodynamic limit near $\lambda=\lambda_c=1-2\alpha$, it is tempting to understand more precisely the scaling near $\lambda=\lambda_c$, that is near $N=2R(1-2\alpha)$. Simple scaling arguments suggest the non trivial behavior occurs when $N-2R(1-2\alpha)$ is of order $R^{1/3}$ (or equivalently $\lambda-\lambda_c$ of order $R^{-2/3}$). Using another time the Geronimo-Case \cite{GeronimoCase}, Borodin-Okounkov \cite{BorodinOkounkov} formula allows to rewrite the partition function in terms of the Fredholm determinant
\begin{equation}
    \tilde{Z}_N(R)=\det_{\ell^2(\mathbb{N})}(1-K_N),
\end{equation}
where $K_N$ is the kernel \eqref{eq:basickern}. We just quote the final result from \cite{Walsh2023}, and provide a sketch of the derivation in Appendix \ref{app:scalingtofred}. If $\alpha<1/8$ then
\begin{equation}\label{eq:cvtw}
    \lim_{R\to \infty} \tilde{Z}_{2R(1-2\alpha)+\sigma[(1-8\alpha)R]^{1/3}}=\det_{L^2(\mathbb{R})}(1-K_{\textrm{Ai},\sigma})
\end{equation}
where $K_{\textrm{Ai},\sigma}$ is the operator with shifted Airy kernel
\begin{align}\label{eq:Airykernel}
    K_{\textrm{Ai},\sigma}(s,s')=\int_\sigma^\infty \textrm{Ai}(s+u)\textrm{Ai}(s'+u) du;
\end{align}
if the index is not an integer we take its integer part. By $\det_{L^2(\mathbb{R})}$ we denote the Fredholm determinant, see Appendix \ref{app:besselfunction}. The rhs in equation \eqref{eq:cvtw} is the celebrated Tracy-Widom \cite{TracyWidom1994} distribution $F_3(\sigma)$. 

On the other hand for $\alpha=1/8$ the behavior is different, and occurs on scales of the order $R^{1/5}$:
\begin{equation}\label{eq:cvtw5}
    \lim_{R\to \infty} \tilde{Z}_{3R/2+\sigma[R/4]^{1/5}}=\det_{L^2(\mathbb{R})}(1-K^{(5)}_{\textrm{Ai},\sigma}).
\end{equation}
This is a higher order Tracy-Widom distribution which was introduced in \cite{LeDoussalMajumdarSchehr2018}. This family has higher order kernels
\begin{align}
    \label{eq:higherairykernel}
    K^{(2m+1)}_{\textrm{Ai},\sigma}(s,s')=\int_\sigma^\infty \textrm{Ai}^{(2m+1)}(s+u)\textrm{Ai}^{(2m+1)}(s'+u) du,
\end{align}
for integer $m\geq 1$, where
\begin{align}\label{eq:higherairyfunction}
    \textrm{Ai}^{(2m+1)}(s)=\int_{\mathbb{R}} \cos\left(ks+\frac{k^{2m+1}}{2m+1}\right)\frac{dk}{2\pi}
\end{align}
is a higher Airy function. The case
$m=1$ corresponds to usual Airy functions, $\textrm{Ai}^{(3)}(s)=\textrm{Ai}(s)$ and gives the Airy kernel \eqref{eq:Airykernel}, while $m=2$ corresponds to \eqref{eq:cvtw5}.
\subsection{Airy scaling near the exterior edges}
\label{sec:airykernelscaling}
Let us zoom close to the exterior edges $X_e=\pm \Upsilon(0)$, where the hydrodynamic density vanishes as square-root and the rescaled edge correlators are expected to converge to the Airy kernel. Due to the left-right symmetry we may just look at the right edge.
The result can be obtained from a saddle-point analysis of the contour integral formulas, or the following simple argument (e.g. \cite{Spohn_lectures,Stephan_free}) based on the hydrodynamic equation \eqref{eq:upsilon2} and the interpretation of its solution $k=k_F$ as the Fermi momentum. Correlations close to a given point $X$ are expected to scale to that of a Fermi sea corresponding to Fermi momentum $k_F$. That is, only real momenta
\begin{equation}
    |k|\leq k_F
\end{equation}
are allowed. Close to the edge $k_F$ is small, and the Taylor expansion of the hydrodynamic equation \eqref{eq:upsilon2} reads
\begin{align}
    \frac{x-x_e}{R}=-\Upsilon''(0) \frac{k^2}{2}
\end{align}
where notice the second derivative is negative, and we have reintroduced the original coordinates. Close to the edge we have a system of free fermions at low density where only momenta 
\begin{equation}
    k^2+\frac{x-x_e}{\tilde{R}}\leq 0
\end{equation}
are allowed, with
\begin{align}
\tilde{R}=- \frac{\Upsilon''(0)}{2}R.    
\end{align}
Semiclassically $k$ may be identified with the operator $\ci \frac{d}{dx}$. In rescaled coordinates $s=\tilde{R}^{1/3}(x-x_e)$, the two point function is expected to project onto the subspace
\begin{equation}
    -\frac{d^2}{ds^2}+s\leq 0.
\end{equation}
As is well known --or can be checked directly-- the eigenfunctions of the operator on the lhs are Airy functions, and the corresponding projection kernel is the Airy kernel \eqref{eq:Airykernel} for $\sigma=0$.

\begin{figure}[htbp]
\includegraphics[width=0.48\textwidth]{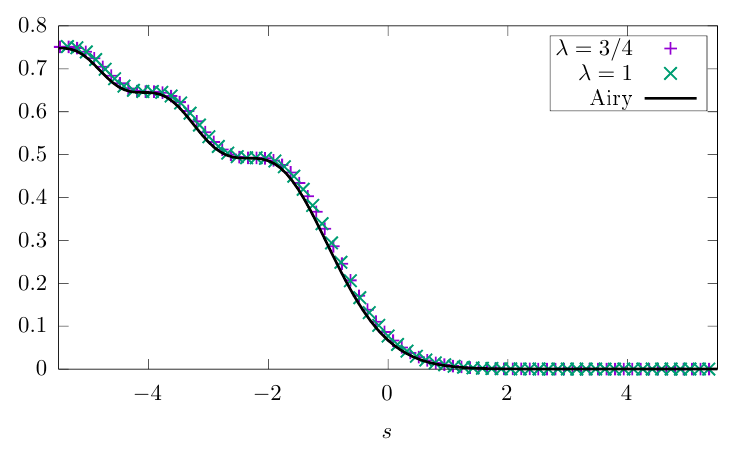}\hfill
\includegraphics[width=0.48\textwidth]{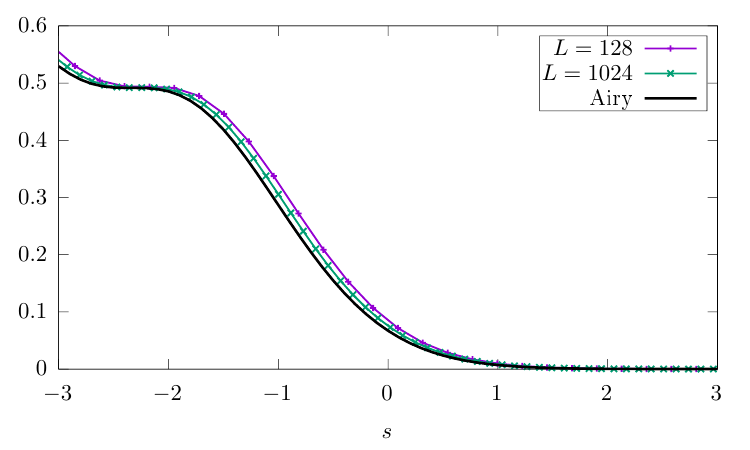}
\caption{Diagonal part of the Airy kernel $K_{Ai}(s,s)$ compared to finite size numerics --the lhs of \eqref{eq:cvAiry}-- for $\alpha=0$. Left: Rescaled density for $L=\lambda R=512$, and $\lambda=3/4,1$ compared to the exact Airy result in the $L\to\infty$ limit. Right: Finite-size effects for $\lambda=3/4$, with $L=128,1024$ compared with the exact limiting result $K_{Ai}(s,s)$.}
\label{fig:tw_convergence}
\end{figure}

Hence we expect
\begin{equation}\label{eq:cvAiry}
    \tilde{R}^{1/3}\braket{c_{x_e+s \tilde{R}^{1/3}}^\dag c_{x_e+s' \tilde{R}^{1/3}}}_{0,R}
    \to K_{\rm Ai,0}(s,s')
\end{equation}
This convergence is illustrated in Figure \ref{fig:tw_convergence} for the density.
Of course, the argument above relies on the fact that $\Upsilon''(0)$ does not vanish, which is indeed always the case for our choice of parameters $0\leq \alpha\leq 1/8$.

However, this can happen if one allows for negative values of $\alpha$, which we temporarily do now. One can show that the top equation in \eqref{eq:upsilondef} still holds with $\alpha=-1/8$ and 
$\lambda \geq 5/4$. In this case the edge limit of the hydrodynamic equation becomes
\begin{align}
    \frac{x-x_{\rm e}}{R}=-\frac{k^4}{8},
\end{align}
and the two point function in appropriately rescaled coordinates becomes the kernel of the projection onto the subspace
\begin{align}
    \frac{d^4}{ds^4}+s\leq 0
\end{align}
whose solution is given by the higher Airy kernel \eqref{eq:higherairykernel} with $m=2$, 
and the actual convergence of the lattice two point function to such kernel has been proved in \cite{Walsh2023} for the simpler domain wall state.

\subsection{Merger and tacnode}
\label{sec:tacnode}
Consider now the center edge(s) illustrated in Figure \ref{fig:densityyzero_plots} for $0\leq \alpha<1/8$. Since the density goes to one in that case, it makes sense to study the hole correlator
\begin{align}\label{eq:holecorr}
    C_{R,L}(x,x')&= (-1)^{x-x'}\braket{c_x c_{x'}^\dag}_{0,R}
\end{align}
where the explicit dependence on $L$ is made more apparent, and an extra phase factor was inserted for later convenience. Notice the diagonal part $C_{R,L}(x,x)$ is simply one minus the density. The case $\lambda>\lambda_c$ is similar \footnote{Formally the particle hole symmetry sends $\alpha\to-\alpha$.} to that described in the previous section. A simple adaptation of the previous arguments leads to the Airy kernels for $0\leq \alpha<1/8$.

The most interesting behavior in our setup occurs at the critical value $\lambda_c=1-2\alpha$ where the two arctic regions merge at $X=0$.
This setup is more complicated because near $X=0$ the two hydrodynamic equations \eqref{eq:upsilon1},\eqref{eq:upsilon2} cannot be seen as independent anymore, one would expect two interacting Airy process rather than a single one. 

To study this limit more precisely, let us use \eqref{eq:rankoneperturbation} and the fact that the generalized Bessel functions form an orthogonal set. The hole correlator is given by
\begin{align}\label{eq:rankoneperturbationbis}
    C_{R,L}(x,x')&= \sum_{p> 0} J^{(\alpha)}_{p+L-x}(R) J^{(\alpha)}_{p+L-x'}(R)+\sum_{j,l\geq 0}a_x(j)(1-K_N)^{-1}_{jl} a_{x'}(l)
\end{align}
where $a_x(j)$ is given by \eqref{eq:axjdef}, and $K_N$ by \eqref{eq:basickern}.

With the notation
\begin{align}
 \tilde{R}=(1-8\alpha)\frac{R}{2},   
\end{align}
we set the scaling
\begin{align}\label{eq:sclimit}
    L&=R(1-2\alpha)+\sigma \tilde{R}^{1/3}\\\label{eq:sclimit2}
    x&=\tilde{R}^{1/3} s\\\label{eq:sclimit3}
    x'&=\tilde{R}^{1/3} s'
\end{align}
for $\alpha< 1/8$, similar but not identical to what was done for the partition function. $\sigma$ quantifies how merged the two regions are, one can think of $\sigma>0$ as the repulsive case, and $\sigma<0$ as the attractive one. $s$ and $s'$ are the rescaled coordinates near the origin. A saddle point analysis shows (see Appendix \ref{app:besselfunction})
\begin{align}
\tilde{R}^{1/3}J_{R(1-2\alpha)+s \tilde{R}^{1/3}}^{(\alpha)}(R)\to \textrm{Ai}(s)
\end{align}
for large $R$, and one can use it to find
\begin{align}
    \tilde{R}^{1/3}C_{R,L}(s\tilde{R}^{1/3},s'\tilde{R}^{1/3}) \to  K_{\rm tac}(s,s'|\sigma)
\end{align}
where $K_{\rm tac}$ is the so-called tacnode kernel \cite{2011_Delvaux,Adler2013,Johansson2013}, under the scaling \eqref{eq:sclimit},\eqref{eq:sclimit2},\eqref{eq:sclimit3}. The $\alpha=0$ case has been proved in \cite{Adler2013,Adler2014}, while the generalization to other values of $\alpha$ would require only very minor modifications to their proof.

The tacnode kernel is more complicated than the Airy one. It is defined in terms of the ancillary function
\begin{align}\label{eq:afunction}
    A_s(u)={\rm Ai}(s+2^{1/3}u+\sigma)-\int_0^\infty {\rm Ai}(u+v+2^{2/3}\sigma) {\rm Ai}(-s+2^{1/3}v+\sigma)dv
\end{align}
as
\begin{align}\label{eq:tackernel}
    K_{\rm tac}(s,s'|\sigma)=
    \int_0^\infty du \textrm{Ai}(u+\sigma-s)\textrm{Ai}(u+\sigma-s')
    +2^{1/3}\int_0^\infty du \int_0^{\infty} dv A_s(u)\left(1-K_{\textrm{Ai},2^{2/3}\sigma}\right)^{-1}(u,v) A_{s'}(v)
\end{align}
where $K_{\textrm{Ai},\sigma}(s,s')=K_{\rm Ai}(s+\sigma,s'+\sigma)$ is the shifted Airy kernel, and the first term in the rhs is $K_{\textrm{Ai},\sigma}(-s,-s')$. Notice the perfect analogy between \eqref{eq:afunction}, \eqref{eq:tackernel} and the discrete formulas \eqref{eq:axjdef},\eqref{eq:rankoneperturbationbis} before taking the limit. The main analytical complication stems from the need to invert (the identity minus) the shifted Airy operator. However this can be done numerically using standard quadrature techniques (e.g. \cite{Bornemann}, see also  \cite{Cafasso} for associated emptiness formation probabilities). While not evident from the formula, it is also symmetric with respect to $s,s'\to -s,-s'$, as well as under the exchange $s\leftrightarrow s'$ as is inherited from the initial problem.

This kernel also contains many others as limiting cases. For example in the infinitely repulsive limit $\sigma\to+\infty$, the operator $\left(1-K_{\textrm{Ai},2^{2/3}\sigma}\right)^{-1}$ acts as $\delta(x-y)$ leading to $K_{\rm Ai}(\sigma-s,\sigma-s')+K_{\rm Ai}(\sigma+s,\sigma+s')$, that is two copies of the Airy kernel, consistent with obvious intuition. A proper attractive scaling limit of an anisotropic tweak of the above tacnode kernel \cite{FerrariVeto} not studied here also leads to the Pearcey kernel \cite{Pearcey}, see \cite{Geudens}, so tacnode can in a sense be seen as a generalization of the previous two.

\begin{figure}[htbp]
\includegraphics[width=0.48\textwidth]{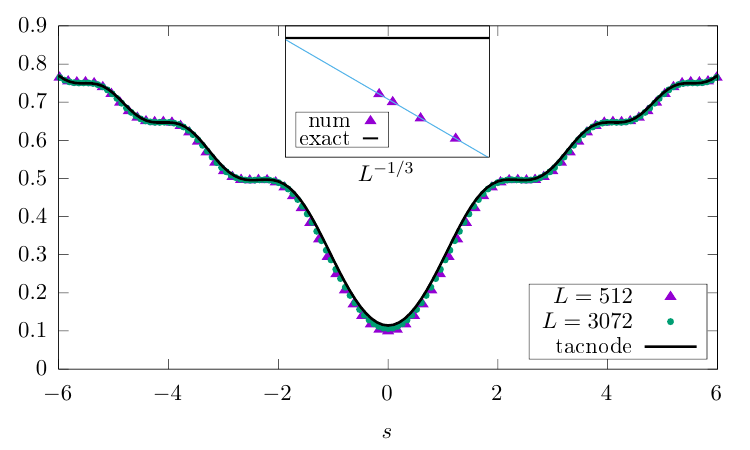}\hfill
\includegraphics[width=0.48\textwidth]{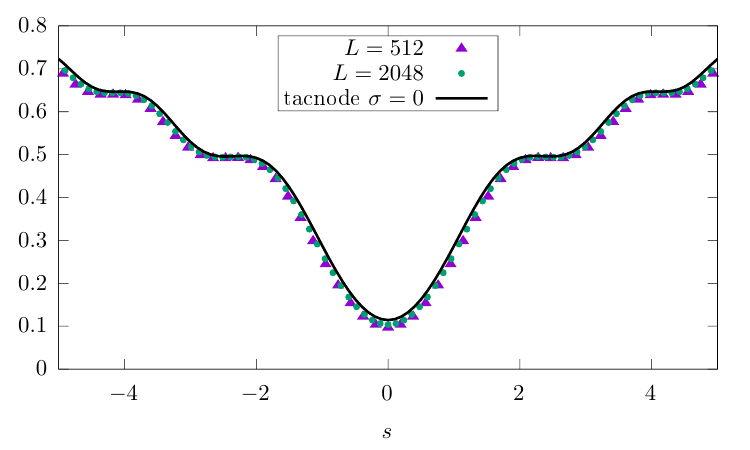}
\includegraphics[width=0.48\textwidth]{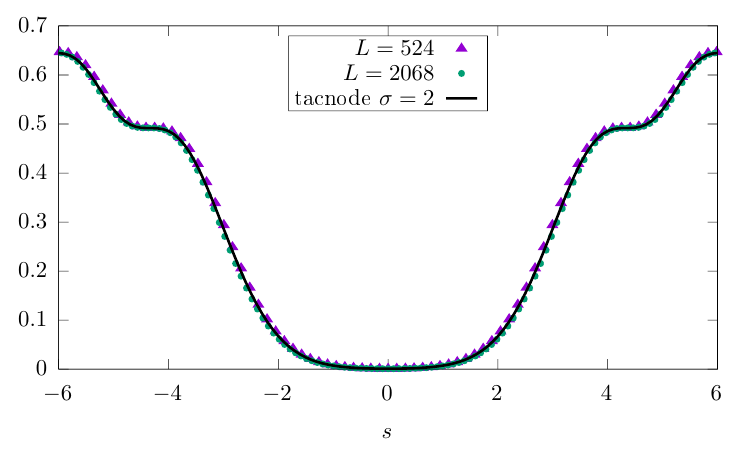}\hfill
\includegraphics[width=0.48\textwidth]{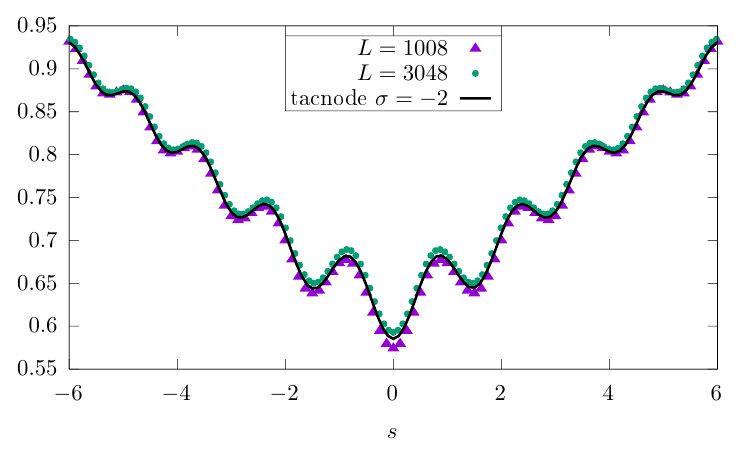}
\caption{Top left: $\alpha=0$. Numerical evaluation of the edge density in the scaling limit \eqref{eq:sclimit} for $L=512,3072$, and comparison with the diagonal part of the tacnode kernel. Remember there are $N=2L+1$ particles. Inset: extrapolation of the value in the middle ($x=0$ or $s=0$) as a straight line in terms of $L^{-1/3}$, with almost perfect agreement with the theoretical result. Top right: same with $\alpha=1/16$ and $L=512,2048$. Bottom left: edge density corresponding to $\sigma=2$, which is very close to two decoupled Airy kernels of Fig.~\ref{fig:tw_convergence} (left). Bottom right: edge density corresponding to $\sigma=-2$ which is very different from the sum of two Airy.}
\label{fig:tac}
\end{figure}

The convergence of the finite $L,R$ result to the limiting kernel is shown in Fig.~\ref{fig:tac}, and illustrates some of those points. Interestingly, the leading correction to the asymptotic behavior is also $R^{1/3}$, as expected from scaling arguments, and as is also the case for usual Airy scaling.

\subsection{Higher merger and higher tacnode}
Here we analyze the particular case $\alpha=1/8$. Above the critical value $\lambda>\lambda_c=3/4$, simple adaptation of the arguments in Section \ref{sec:airykernelscaling} lead to a $m=2$ higher Airy kernel close to the two center edges at $X=\pm \Upsilon(\pi)=\lambda-\lambda_c$. 

\begin{figure}[htbp]
\includegraphics[width=0.48\textwidth]{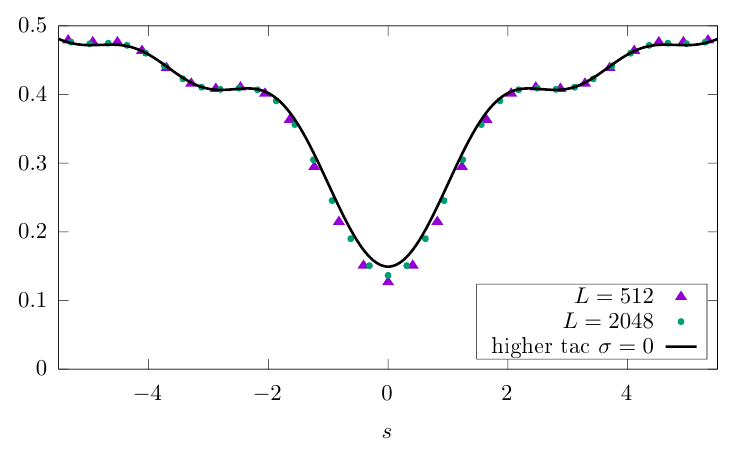}\hfill
\includegraphics[width=0.48\textwidth]{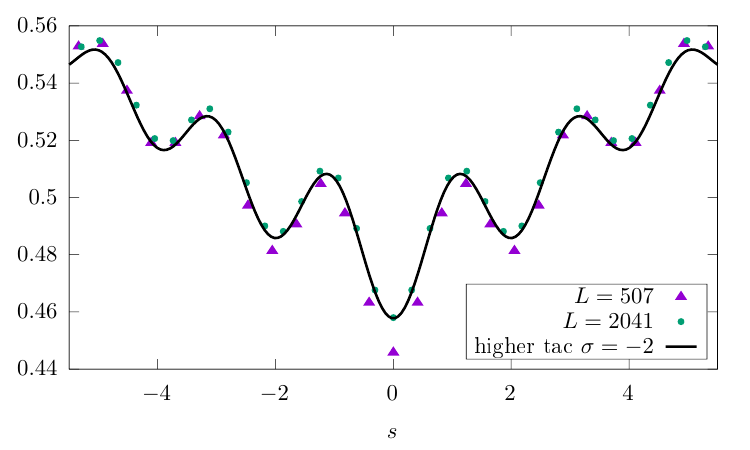}
\caption{Higher critical value $\alpha=1/8$, $\lambda=\lambda_c=3/4$. Left: numerical evaluation of the edge density in the scaling limit \eqref{eq:higherscaling} for $L=512,2048$ and $\sigma=0$. Right: numerical edge density in the same scaling limit for 
$L=507,2041$ but $\sigma=-2$. In both cases the finite size values are reasonably close to the limiting kernel \eqref{eq:tac5} for large $L$.}
\label{fig:highertacnode}
\end{figure}

The case $\lambda=\lambda_c$ is most interesting because the two fluctuating regions merge, combined with the fact that the (hole) density vanishes as a fourth root, see \eqref{eq:fourthroot}. This means one gets a higher tacnode \footnote{In the language of algebraic geometry, algebraic curves of the form $X^2=Y^{j+1}$ provide examples of $A_j$-curve singularities at $(0,0)$. The cusp corresponds to $j=2$, the tacnode corresponds to $j=3$, while our higher tacnode corresponds to $j=7$.} and one expects different universal behavior. Using the exact formula  \eqref{eq:rankoneperturbation} with now
\begin{align}
\tilde{R}=\frac{R}{8}
\end{align}
and the scaling
\begin{align}\label{eq:higherscaling}
    L&=\frac{3R}{4}+\sigma \tilde{R},\\
    x&=\tilde{R}^{1/5}s\\
    x'&=\tilde{R}^{1/5}s'
\end{align}
we predict the new limiting kernel
\begin{align}
    \tilde{R}^{1/5}C_{R,L}(s \tilde{R}^{1/5},s' \tilde{R}^{1/5}) \to K_{\rm tac}^{(5)}(s,s'|\sigma),
\end{align}
given by
\begin{align}\label{eq:tac5}
    K_{\rm tac}^{(5)}(s,s'|\sigma)=K_{\rm Ai}^{(5)}(\sigma-s,\sigma-s')+2^{1/5}\int_0^\infty dx \int_0^{\infty} dy A^{(5)}_s(x)\left(1-K^{(5)}_{\textrm{Ai},2^{4/5}\sigma}\right)^{-1}(x,y) A^{(5)}_{s'}(y),
\end{align}
where
\begin{align}\label{eq:a5}
    A^{(5)}_s(x)={\rm Ai}^{(5)}(s+2^{1/5}x+\sigma)-\int_0^\infty {\rm Ai}^{(5)}(x+y+2^{4/5}\sigma) {\rm Ai}^{(5)}(-s+2^{1/5}y+\sigma)dy.
\end{align}
In Fig.~\ref{fig:highertacnode}, we compare the finite size rescaled density to our predicted kernel. Finite-size effects are bigger, as expected from $R^{1/5}$ scaling, but the overall agreement is still very good. Notice also that under this integral form the structure is very similar to that of regular tacnode, however we expect alternative forms to look different, as is discussed in \cite{LeDoussalMajumdarSchehr2018}, for the higher Airy process. 
\subsection{A simpler quench problem}
Let us investigate the center edge behavior of a quench problem which has been studied in \cite{ScopaCalabreseDubail2022}. The initial state and the Hamiltonian are the same as in our setup, with $\alpha=0$. The only difference is that time evolution is governed by the Schrödinger equation
\begin{align}
\braket{N|e^{\ci H t}c_x^\dag c_{x'} e^{-\ci H t}|N}&=\ci^{x-x'}\sum_{j=-L}^{L}J_{x-j}(t)J_{x'-j}(t)
\end{align}
where the case $x=x'$ gives the density $\rho(x,t)$. 
The interesting behavior occurs for $L/t$ close to one. In this case the hydrodynamic density profile is given for large $x,t$ by \cite{ScopaCalabreseDubail2022}
\begin{equation}
    \rho(x,t)\sim \frac{1}{\pi}\arccos \left(\left|\frac{x}{t}\right|-1\right)
\end{equation}
so looks exactly like the profiles studied here in the tacnode setup. However the hole correlator can be written as
\begin{align}
    \tilde{C}_{t,L}(x,x')&=\braket{N|e^{\ci H t}c_x c_{x'}^\dag e^{-\ci H t}|N}\\
    &=\ci^{x-x'}\sum_{j>0}\left(J_{j+L+x}(t)J_{j+L+x'}(t)+(-1)^{x-x'}J_{j+L-x}(t)J_{j+L-x'}(t)\right)
\end{align}
which is the sum of two semi-infinite domain wall correlators (e.g.  \cite{EislerRacz2013,Stephan_free}).
Similar to previous sections we zoom in close to the center, and
after saddle point analysis we obtain
\begin{equation}
  \left(\frac{t}{2}\right)^{1/3}  \tilde{C}_{t,t+\sigma \left(\frac{t}{2}\right)^{1/3}}\left(s\left(\frac{t}{2}\right)^{1/3},s'\left(\frac{t}{2}\right)^{1/3}\right)\sim   K_{\rm Airy}(\sigma+s,\sigma+s')+(-1)^{x-x'} K_{\rm Airy}(\sigma-s,\sigma-s').
\end{equation}
This behavior is identical to that studied in \cite{Walsh2024} for a different model. An important difference with tacnode is that it is, up to a fast oscillating phase, the sum of two independent Airy kernels, one determined from the left moving free particles, the other from the right moving ones. Therefore, a hydrodynamic density profile with merger of fluctuating region does not guarantee the appearance of a tacnode process, for which a nontrivial statistical interaction between two Airy processes is necessary.

\section{Conclusion}
\label{sec:conclusion}
In this paper, our starting point was a nearest neighbor free fermionic model evolving in imaginary time from a double domain wall state. It is known to give rise to limit shapes, and the parameters can be tuned in such a way that two fluctuating regions touch as emphasized in \cite{Pallister_2022}. The hydrodynamic density has a square-root singularity at the touching point, and one expects universal behavior governed by a tacnode process in its vicinity, with scaling $R^{1/3}$, where $R$ is a characteristic size of the fluctuating regions. This has been shown in a different self-avoiding random walk model \cite{Adler2013} which however gives rise to the same correlation functions (in probabilistic parlance the same determinantal point process). 

We then added a next-nearest neighbor term, and tuned it in such a way to obtain a quartic behavior of the dispersion at low momenta, which turns into a fourth-root singularity for the density at the touching point. In the vicinity of this touching point we found a new (higher-order) tacnode process which we determined analytically. As emphasized in the text, the two crucial ingredients to get such a higher-order tacnode are (1) a merging of fluctuating regions (2) higher-order vanishing for the density profile. We expect this higher-order tacnode to appear for a much broader class of inhomogeneous states, even though deriving it from lattice computations--despite the free fermionic nature of the model--would presumably pose considerable technical difficulties.

There are several ways in which our findings can be generalized. The most obvious one is to consider more general dispersion relations of the form $\varepsilon(k)=\sum_{p\geq 1} \alpha_p \cos (pk)$. In this case formulas \eqref{eq:rankoneperturbation}, \eqref{eq:axjdef}, \eqref{eq:basickern} still hold, albeit with $J_n^{(\alpha)}$ replaced by 
\begin{align}
    J_n^{(\alpha_1,\alpha_2,\ldots)}(R)
    &=\int_{-\pi}^{\pi} \frac{dk}{2\pi} e^{-\ci k n} e^{\ci R \sum_{p\geq 1} (-1)^{p+1} \alpha_p  \sin (pk)}.
\end{align}
Known classical lattice models fit into this framework. 
For example (see e.g. \cite{Allegra_2016}), the dispersion relation for classical dimers on the honeycomb lattice with some fugacity $u$ is $\varepsilon(k)=\frac{1}{2}\log\left[(1+u e^{\ci k})(1+u e^{-\ci k})\right]$. It is also possible to tune the parameters $\alpha_p$ to get even higher order scaling, as was done systematically in \cite{Walsh2023} without merging. In this case the new limiting tacnode kernel of order $2m+1$ will be given by \eqref{eq:higherairykernel}, \eqref{eq:tac5}, \eqref{eq:a5}, with $1/5$ replaced by $1/(2m+1)$, $4/5$ replaced by $2m/(2m+1)$, and $^{(5)}$ replaced by $^{(2m+1)}$. We did not discuss this further in the paper because their structure is very similar, instead focusing on the simplest higher one.

An important distinction between higher kernels and normal ones lies in the fact that higher kernels do not have extended versions at unequal imaginary time \cite{Walsh2023}, even though the finite size correlators can be computed using similar methods. This is related to the positivity issue for the hydrodynamic density profile discussed in \cite{Bocini2021}, and the appearance of crazy regions away from $y=0$. We expect the same issue in the tacnode setup, as is illustrated numerically in the bottom part of Fig.~\ref{fig:densitiesalpha}, where the crazy region nearly touches the tacnode point precisely at the critical value $\alpha=1/8,\lambda=\lambda_c$.

As explained in \cite{Bocini2021} the positivity issues for $\alpha>0$ and $y\neq 0$ can be cured by considering lower density initial states. For example in our setup, we would expect the initial state $\prod_{|j|\leq 2L}c_{2j}^\dag|\ket{0}$ with half density in the center to yield a positive density profile at all times, and it should be possible to attack this problem with techniques based on orthogonal polynomials as well. 

The tacnode process is more general than the Airy process, but it is not the most general one \cite{Johansson_review}. For example one can engineer similar touching arctic curves but which do not have the same curvature at the touching point. The corresponding anisotropic tacnode process has been studied in \cite{FerrariVeto} and has a similar analytic expression. 
In fermionic language it is also possible to engineer such a setup by considering non symmetric hoppings, and study higher versions in a systematic way. Finally, one could do the same to access the discrete tacnode kernel \cite{Discretetacnode}, which has the tacnode kernel as a limit.

A more difficult question deals with the effect of (sufficiently short-range) interactions, where universality is expected close to the edge by a simple dilution argument (e.g. \cite{Stephan_free}). Analytical progress has been slow but steady \cite{ColomoPronkocurve,ColomoPronkoZinn,deGier2021,Colomo2024,Yeh_2022,polytropic}. A possibility would be to check numerically for the emergence of the tacnode kernel using Monte Carlo simulations in the six-vertex model similar to that in \cite{Viti_2023,Prahofer_2024}. However this is presumably not so easy, especially since (in the repulsive case $\sigma\geq 0$) the tacnode and (two copies of) Airy are numerically very close.

Perhaps the most important question deals with the realization of our process as a quantum ground state, interacting or not, or through some natural out of equilibrium setup, interacting or not.
To check for Airy scaling in interacting systems one can use the trick of \cite{Stephan_free,Eislergradient}, which studies the ground state of an inhomogeneous spin chain where it is expected to appear. Such quantum ground state are also perhaps easier to access experimentally. More simply one can accurately access the ground state for large systems using DMRG simulations. However, it is not immediately clear how to engineer an interacting ground state with tacnode, due to the anomalous terms appearing already in the free case \eqref{eq:recursion_phi}, and which seem to prevent a simple interpretation of the $\phi_k(x)$ as single particle wave functions stemming from a simple (hermitian) quantum Hamiltonian. We are however not able to exclude that this is ultimately possible, and leave this as an open problem.

\vspace{1.5cm}
\textbf{Acknowledgments.}
We are grateful to Sasha Abanov, Saverio Bocini, Jérémie Bouttier, Filippo Colomo and Sofia Tarricone for useful discussions, and thank Filippo Colomo for careful reading of the manuscript.

\clearpage

\appendix 

\section[\qquad \qquad \qquad Orthogonal polynomials on the unit circle]{Orthogonal polynomials on the unit circle}
\label{app:orthopoly}
In this appendix, we gather several results regarding orthogonal polynomials on the unit circle $S^1=\{z\in \mathbb{C},|z|=1\}$ which are used extensively in the main text. Most of these concepts are classical and can be found e.g. in \cite{Szegobook,simon2005orthogonal,Assche_2020}.

Consider some weight function $w(z)$. We wish to find polynomials $P_n(z)$ of exact degree $n$ which are orthonormal with respect to this weight on $S^1$:
\begin{eqnarray}
\braket{P_n|P_l}&=\oint_{|z|=1} \frac{dz}{2\ci \pi z}w(z)\overline{P_n(z)}P_l(z) \\
    &=\int_{-\pi}^{\pi}\frac{d\theta}{2\pi}w(e^{\ci \theta})\overline{P_n(e^{\ci \theta})}P_l(e^{\ci \theta})\\
    &=\delta_{nl}
\end{eqnarray}
We ask for the coefficient of the leading term to be positive, that is
\begin{equation}
    P_n(z)=\kappa_n z^n+\ldots\qquad,\qquad \kappa_n>0
\end{equation}
To simplify the exposition we also assume $w(e^{\ci \theta})\geq 0$ and $w(e^{\ci \theta})=w(e^{-\ci \theta})$. In particular this implies that $P_n$ has real coefficients, so $\overline{P_n(e^{\ci \theta})}=P_n(e^{-\ci \theta})$.

The $(P_k)_{0\leq k\leq n}$ form a basis of the space of polynomials of degree $\leq n$, hence  $P_n$ is orthogonal to any polynomial of degree $n-1$ or less, so in particular all monomials $z^{n-1},\ldots,z,z^0$. In the following an important role will be played by the reverse polynomials
\begin{eqnarray}
    P_n^*(z)=z^n P_n(1/z).
\end{eqnarray}
As follows from the definition, $P_n^*$ is orthogonal to all monomials $z,z^1,\ldots z^{n}$, so all polynomials of degree $\leq n$ that vanish at the origin $z=0$. It is (up to an overall factor) the unique such polynomial.
\subsection[\qquad \qquad \qquad Szeg\"o recurrences]{Szeg\"o recurrences}
Introduce 
\begin{eqnarray}
    \rho_n=\frac{\kappa_n}{\kappa_{n+1}}.
\end{eqnarray}
It is easy to check that $\rho_n P_{n+1}(z)-zP_n(z)$ is orthogonal to $z^{n},\ldots,z$, so we have the Szeg\H{o} recurrence
\begin{equation}\label{eq:Szegorecursion}
    \rho_n P_{n+1}(z)=z P_n(z)-\alpha_n P_n^*(z).
\end{equation}
The $\alpha_n$ are sometimes called Verblunsky coefficients. There is also a three term recurrence not involving the reverse polynomials:
\begin{equation}\label{eq:threeterms}
    \rho_n P_{n+1}(z)=\left(\frac{\alpha_n}{\alpha_{n-1}}+z\right)P_n(z) -\frac{\alpha_n}{\alpha_{n-1}}\rho_{n-1} z P_{n-1}(z).
\end{equation}
\subsection[\qquad \qquad \qquad Christoffel-Darboux kernel]{Christoffel-Darboux kernel}
Introduce
\begin{equation}\label{eq:cd_def}
    K_n(z,\zeta)=\sum_{l=0}^n P_l(z)\overline{P_l(\zeta)}.
\end{equation}
It satisfies
\begin{equation}
    \oint \frac{dz w(z)}{2\ci \pi z}P(z) \overline{K_n(z,\zeta)}=P(\zeta)
\end{equation}
for any polynomial $P$ of degree $n$ or less. Otherwise it projects onto the the subspace of polynomials with degree $\leq n$. We also have the formula
\begin{align}\label{eq:cd_formula}
    K_n(z,\zeta)&=\frac{P_n^*(z)\overline{P_n^*(\zeta)}-\bar{\zeta}z P_n(z)\overline{P_n(\zeta)}}{1-\bar{\zeta}z}\\\label{eq:cd_formulabis}
    &=\frac{P_{n+1}^*(z)\overline{P_{n+1}^*(\zeta)}-P_{n+1}(z)\overline{P_{n+1}(\zeta)}}{1-\bar{\zeta}z}.
\end{align}
which both go under the name Christoffel-Darboux formula. 
The particular case $\zeta=0$,
\begin{equation}\label{eq:cdparticular}
    \kappa_n P_n^*(z)=- \sum_{l=0}^n \kappa_l \alpha_{l-1}P_l(z),
\end{equation}
allows to express $P_n^*$ in terms of the $P_n$.
\subsection[\qquad \qquad \qquad Relation to Toeplitz matrices]{Relation to Toeplitz matrices}\label{app:poly_toeplitz}
Consider the (real symmetric) matrix with elements
\begin{equation}
    A[w]_{jl}=\oint \frac{dz\, w(z)}{2\ci \pi z^{j-l+1}}
\end{equation}
built from the weight $w$, which depends only on $j-l$ (Toeplitz matrix). We denote by
\begin{equation}
    D_n[w]=\det_{0\leq j,l\leq n-1}\left(A[w]_{jl}\right)
\end{equation}
the corresponding Toeplitz determinant. All previously encountered coefficients may be expressed in terms of such determinants:
\begin{eqnarray}
    \kappa_n^2&=&\frac{D_n}{D_{n+1}} \\\label{eq:rhon}
    \rho_n^2&=&1-\alpha_n^2=\frac{D_{n+2}D_n}{D_{n+1}^2}\\
    \alpha_n&=&(-1)^n \frac{D_{n+1}[\tilde{w}]}{D_{n+1}[w]}
\end{eqnarray}
where $\tilde{w}(z)=\frac{1}{z}w(z)$. 

The inverse of a Toeplitz matrix can be expressed in terms of the Christoffel-Darboux kernel
\begin{equation}\label{eq:toeplitzinverse}
    \left(A[w]^{-1}\right)_{jl}=\oint_{|z|=1} \frac{dz}{2\ci \pi z^{l+1}} \oint_{|\zeta|=1} \frac{d\zeta}{2\ci \pi \zeta^{-j+1}}K_{n-1}(z,\zeta)
\end{equation}
with indices $j,l\in \{0,\ldots,n-1\}$.
The orthogonal polynomials themselves may be expressed as Toeplitz determinants
\begin{equation}\label{eq:polydet}
    P_n(z)=\frac{D_n[(z-1/\zeta)w(\zeta)]}{\sqrt{D_n[w]D_{n+1}[w]}}=\frac{D_n[(z-\zeta)w(\zeta)]}{\sqrt{D_n[w]D_{n+1}[w]}}
\end{equation}
and even the Christoffel-Darboux kernel itself can be expressed as a ratio of Toeplitz determinants
\begin{align}\label{eq:cdkerneldet}
    \sum_{l=0}^n P_l(z)P_l(1/z')&=\left(\frac{z}{z'}\right)^n \frac{D_n[(1-\zeta/z)(1-z'/\zeta)w(\zeta)]}{D_{n+1}[w(\zeta)]}.
\end{align}
\subsection[\qquad \qquad \qquad The Geronimo-Case Borodin-Okounkov  formula]{The Geronimo-Case Borodin-Okounkov formula}
\label{app:gcboformula}
This formula was discovered in \cite{GeronimoCase}, see also \cite{BorodinOkounkov}. Write $w(z)=e^{g(z)}$, where
\begin{equation}
    g(z)=g_+(z)+g_-(z)\qquad,\qquad g_{\pm}(z)=\sum_{k\geq 1} g_{\pm k}z^k
\end{equation}
Define also
\begin{equation}
    g^*(z)=g^-(-z)-g^+(-z)
\end{equation}
Modulo assumptions on the regularity of $g$, we have the identity
\begin{equation}\label{eq:gc}
    D_n[e^g]=\exp\left(\sum_{k\geq 1} k g_k g_{-k}\right)\det_{\ell^2(\mathbb{Z}_{\geq n})}(1-\mathcal{K})
\end{equation}
where $\det_{\ell^2(\mathbb{Z}_{\geq n})}(1-\mathcal{K})$ denotes the Fredholm determinant of the operator acting on sequences in $\ell^2(\mathbb{Z}_{\geq n})$ with kernel
\begin{equation}\label{eq:gckern}
    \mathcal{K}_{jl}=\oint \frac{dz}{2\ci \pi z^{j+1}}\oint \frac{d\zeta}{2\ci \pi \zeta^{-l}}\frac{e^{g^*(z)-g^*(\zeta)}}{z-\zeta},
\end{equation}
\begin{equation}
    \det_{\ell^2(\mathbb{Z}_{\geq n})}(1-\mathcal{K})=\exp\left(-\sum_{p\geq 1} \frac{1}{p}\sum_{i_1=n}^\infty \ldots \sum_{i_p=n}^\infty \mathcal{K}_{i_1 i_2}\ldots \mathcal{K}_{i_p i_1}\right)
\end{equation}
A simple consequence of (\ref{eq:gc}) is
\begin{equation} \label{eq:szegothm}
   \lim_{n\to\infty} D_n[e^g]= \exp\left(\sum_{k\geq 1} k g_k g_{-k}\right)
\end{equation}
which is known as strong Szeg\H{o} limit theorem \cite{Szego}. One can also combine (\ref{eq:polydet}) and (\ref{eq:gc}) to express orthogonal polynomials as Fredholm determinants, and similarly for the Christoffel-Darboux kernel.
\subsection[\qquad \qquad \qquad Underlying integrability]{Underlying integrability}
It is possible to express the derivative of $P_n$ in terms of $P_n$ and $P_n^*$. Indeed expressing $P_n'$ as a linear combination of the $(P_k)_{0\leq k\leq n-1}$, we obtain after some algebra (e.g. \cite{Assche_2020})
\begin{equation}\label{eq:dpdz}
   \frac{d P_n(z)}{dz}=a_n(z)P_n(z)-b_n P_n^*(z)
\end{equation}
with
\begin{eqnarray}\label{eq:an}
    a_n(z)&=& \frac{n}{z}+\oint_{|\zeta|=1} \frac{d\zeta e^{g(\zeta)}}{2\ci \pi \zeta}\overline{P_n(\zeta)}P_n(\zeta)\frac{g'(\zeta)-g'(z)}{1-z/\zeta}  \\\label{eq:bn}
    b_n(z)&=&-\frac{n \alpha_{n-1}}{z}+\oint_{|\zeta|=1} \frac{d\zeta e^{g(\zeta)}}{2\ci \pi \zeta} \overline{P_n^*(\zeta)}P_n(\zeta)\frac{g'(\zeta)-g'(z)}{1-z/\zeta}
\end{eqnarray}
A similar formula holds for $\frac{d P_n^*(z)}{dz}$. Write $X_n=\left(\begin{array}{c}P_n\\P_n^*\end{array}\right)$. From the Szeg\H{o} recursion
\begin{equation}
    X_{n+1}=L_n X_n\qquad,\qquad L_n=\left(\begin{array}{cc}z&-\alpha_n\\-\alpha_n z&1\end{array}\right),
\end{equation}
and
\begin{equation}
    \frac{d X_n}{dz}=M_n X_n\qquad,\qquad M_n=\left(\begin{array}{cc}a_n(z)&-b_n(z)\\b_n(1/z)/z^2&\frac{n}{z}-a_n(1/z)/z^2\end{array}\right).
\end{equation}
$L_n,M_n$ form a Lax pair. The compatibility equation for this pair reads
\begin{equation}\label{eq:compatibility}
    \frac{d L_n}{dz}=M_{n+1}L_n-L_n M_n.
\end{equation}

\section[\qquad \qquad \qquad Asymptotics for a generalized Bessel weight]{Asymptotics for a generalized Bessel weight} \label{app:asymptotics_Bessel}
In this appendix, we study the orthogonal polynomials corresponding to the specific weight
\begin{equation}
    w(z)=e^{g(z)}=\exp\left(R\left[z+\frac{1}{z}+\alpha\left(z^2+\frac{1}{z^2}\right)\right]\right)
\end{equation}
which is directly related to our free fermions problem, since $w(e^{\ci \theta})=e^{2 R \varepsilon(\theta)}$, see \eqref{eq:dispersion}. Some properties of the polynomials have already been studied in \cite{ClioCresswell_1999,ChouteauTarricone}.
\subsection[\qquad\qquad\qquad Lax Pair and recursion relations]{Lax Pair and recursion relations}
For this weight
\begin{eqnarray}\nonumber
    \frac{1}{R}\frac{g'(\zeta)-g'(z)}{1-z/\zeta}=\frac{1}{z^2}+\frac{1}{z\zeta}+2\alpha\left(\frac{1}{z^3}+\frac{1}{z^2\zeta}+\frac{1}{z\zeta^2}+\zeta\right)
\end{eqnarray}
so equations (\ref{eq:an}),(\ref{eq:bn}) become
\begin{eqnarray} \label{eq:bess_an}
    a_n(z)&=&\frac{n}{z} +R\left[\frac{1}{z^2}+\frac{2\alpha}{z^3}+2\alpha \braket{P_n|\zeta P_n}+\frac{z+2\alpha}{z^2}\braket{\zeta P_n |P_n}+\frac{2\alpha}{z}\braket{\zeta^2 P_n |P_n}\right]\\ \label{eq:bess_bn}
    b_n(z)&=&-\frac{n\alpha_{n-1}}{z}+R\left[\frac{z+2\alpha}{z^3}\braket{P_n^*|P_n}+2\alpha  \braket{P_n^*|\zeta P_n}+\frac{z+2\alpha}{z^2}\braket{\zeta P_n^* |P_n}+\frac{2\alpha }{z}\braket{\zeta^2 P_n^* |P_n}\right]
\end{eqnarray}
where it is understood that $P_n=P_n(\zeta)$. All scalar products can be computed recursively from the Szeg\H{o} recursion (\ref{eq:Szegorecursion}) as well as (\ref{eq:cdparticular}):
\begin{eqnarray}\label{eq:alphandef}
\braket{P_n^*|P_n}&=&-\alpha_{n-1}\\
\braket{P_n^*|\zeta P_n}&=&\alpha_n\\
\braket{\zeta P_n|P_n}&=&-\alpha_{n-1}\alpha_n=\braket{P_n|\zeta P_n}\\
\braket{\zeta P_n^*|P_n}&=&\alpha_{n-1}^2\left(\alpha_n+\alpha_{n-2}\right)-\alpha_{n-2}\\
\braket{\zeta^2 P_n|P_n}&=&\beta_n=\alpha_n \left[\alpha_{n-1}^2(\alpha_n+\alpha_{n-2})-\alpha_{n-2}\right]-(1-\alpha_n^2)\alpha_{n-1}\alpha_{n+1}\\
\braket{\zeta^2 P_n^*|P_n}&=&\gamma_{n}
\end{eqnarray}
Using the compatibility equation for the Lax pair we obtain
\begin{equation}
    \gamma_n/R=\alpha_{n-3}(\alpha_{n-2}^2-1)+\alpha_{n-1}\left(\alpha_{n-2}^2-\beta_{n-1}-\beta_n\right)
\end{equation}
as well as the recursion relation
\begin{equation} \label{eq:pIIrecursion}
    \frac{(n+1)\alpha_n}{R \rho_n^2}+\alpha_{n-1}+\alpha_{n+1}+2\alpha\!\left(\alpha_{n-2}\rho_{n-1}^2-\alpha_n[\alpha_{n-1}+\alpha_{n+1}]^2+\alpha_{n+2}\rho_{n+1}^2\right)=0
\end{equation}
where recall $\rho_n^2=1-\alpha_n^2$. For consistency with \eqref{eq:alphandef} we use the convention $\alpha_{-1}=-1$. This recursion is used in the main text (see \eqref{eq:discretePIImaintext}), with $\alpha_n=(-1)^n u_{n+1}$.

Plugging all this in \eqref{eq:an},\eqref{eq:bn} yields 
\begin{align}\label{eq:an_tot}
\frac{a_n(z)}{R}&=-2\alpha \alpha_{n-1}\alpha_n +\frac{n/R-\alpha_{n-1}\alpha_n+2\alpha \beta_n}{z}+\frac{1-2\alpha \, \alpha_{n-1}\alpha_n}{z^2}+\frac{2\alpha}{z^3}\\
\label{eq:bn_tot}
    \frac{b_n(z)}{R}&=2\alpha \alpha_n+\frac{[\alpha_{n-1}^2(\alpha_n+\alpha_{n-2})-\alpha_{n-2}]-n\alpha_{n-1}/R+2\alpha \gamma_n}{z}+\frac{2\alpha[\alpha_{n-1}^2(\alpha_n+\alpha_{n-2})-\alpha_{n-2}]-\alpha_{n-1}}{z^2}-\frac{2\alpha \alpha_{n-1}}{z^3}.
\end{align}
Let us now recover the (almost) eigenvalue equation given in the main text, in the case $\alpha=0$ for simplicity. In this case we have
\begin{align}
    a_n(z)&=\frac{n-R\alpha_{n-1}\alpha_n}{z}+\frac{R}{z^2}\\
    b_n(z)&=R\left(\frac{\alpha_n }{z}-\frac{\alpha_{n-1}}{z^2}\right)
\end{align}
Recalling \eqref{eq:phik}
\begin{equation}
    \phi_n(x)=\oint \frac{dz}{2\ci \pi z^{x+L+1}}P_n(z) e^{\frac{R}{2}(z+z^{-1})}
\end{equation}
and integrating by parts, we obtain
\begin{align}
    (x+L)\phi_n(x)&=\oint \frac{dz}{2\ci\pi z^{L+x}}Q_n'(z)\\
   &= \oint \frac{dz}{2\ci\pi z^{L+x}}\left[\frac{n-R\alpha_{n-1}\alpha_n}{z}+\frac{R}{2}\left(1+\frac{1}{z^2}\right)\right]Q_n(z)-R \oint \frac{dz}{2\ci\pi z^{L+x}}\left[\frac{\alpha_n}{z}-\frac{\alpha_{n-1}}{z^2}\right] Q_n^*(z)\\
   &=\oint \frac{dz}{2\ci\pi z^{L+x+1}}\left[n-R\alpha_{n-1}\alpha_n+\frac{R}{2}\left(z+\frac{1}{z}\right)\right]Q_n(z)-R \oint \frac{dz z^n}{2\ci\pi z^{L+x+1}}\left[\alpha_n-\frac{\alpha_{n-1}}{z}\right] Q_n(1/z)\\
   &=\oint \frac{dz}{2\ci\pi z^{L+x+1}}\left[n-R\alpha_{n-1}\alpha_n+\frac{R}{2}\left(z+\frac{1}{z}\right)\right]Q_n(z)-R \oint \frac{dz z^{L+x}}{2\ci\pi z^{n+1}}\left[\alpha_n-\alpha_{n-1}z\right] Q_n(z)
\end{align}
Back in terms of the $\phi_n$, we obtain
\begin{align}\label{eq:phik_recurrence}
    \frac{(x+L)\phi_n(x)}{R}=\left(\frac{n}{R}-\alpha_{n-1}\alpha_n\right)\phi_n(x)+\frac{\phi_{n}(x+1)+\phi_{n}(x-1)}{2}-\alpha_n \phi_n(n-2L-x)+\alpha_{n-1}\phi_n(n-2    L-x-1)
\end{align}
Replacing $\alpha_n$ by $(-1)^n u_{n+1}$ we obtain the result \eqref{eq:recursion_phi} advertised in the main text.

\subsection[\qquad\qquad\qquad Ratio of consecutive polynomials]{Ratio of consecutive polynomials}
Here our goal is to derive an asymptotic formula for the ratio $P_{N+1}(z)/P_N(z)$ in the hydrodynamic limit ($R,N\to\infty$ with $\lambda=N/(2R)$ fixed), and with $|z|>1$. We assume $\alpha\in [0,1/8]$.

As in the main text we find it more convenient to work with $u_n=(-1)^{n-1}\alpha_{n-1}$. The first step is to assume $u_N\to u$ in the above limit. From \eqref{eq:pIIrecursion} we find that $u=u(\lambda,\alpha)$ is the solution in $[0,1]$ to
\begin{align}\label{eq:someequation}
    \frac{\lambda u}{1-u^2}=u-2\alpha u (1-3u^2).
\end{align}
This result is used in the main text to obtain the hydrodynamic limit of the partition function, but it is also useful to study ratios of polynomials. Indeed the three term recurrence \eqref{eq:threeterms} can be rewritten as
\begin{align}
    \rho_n \frac{P_{n+1}(z)}{P_n(z)}=\left(z-\frac{u_{n+1}}{u_n}\right)+\frac{u_{n+1}}{u_n}\rho_{n-1}z\frac{P_{n-1}(z)}{P_n(z)}.
\end{align}
This means we expect the ratio convergence $P_{n+1}(z)/P_n(z)\to r(z)$ which satisfies the second degree equation
\begin{equation}
    \sqrt{1-u^2}r(z)^2=(z-1)r(z)+\sqrt{1-u^2}z.
\end{equation}
Among the possible solutions, we select 
\begin{equation} \label{eq:correcchoice}
r(z) = \frac{z - 1 + \sqrt{(z + 1)^2 - 4u^2 z}}{2\sqrt{1 - u^2}}
\end{equation}
by a continuity argument which we explain now. For $\lambda\geq \lambda_c$ the solution to \eqref{eq:someequation} is simply $u=0$, which would imply $r(z)=z$ (the other choice of root would give $r(z)=-1$). The fact that the former choice is the correct one is strongly suggested by \eqref{eq:polydet} which implies
\begin{align}
    \frac{P_{n+1}(z)}{P_n(z)}=z \frac{D_{n+1}[(1-1/(\zeta z))w(\zeta)]}{D_n[(1-1/(\zeta z))w(\zeta)]} \sqrt{\frac{D_n[w(\zeta)]}{D_{n+2}[w(\zeta)]}}
\end{align}
and applying the Szeg\H{o} limit theorem yields
\begin{align}
    \lim_{n\to \infty} \frac{P_{n+1}(z)}{P_n(z)}=z
\end{align}
when $|z|>1$. Since the result at $\lambda=\infty$ is expected to hold for any $\lambda>1$ --or can be shown more precisely by making use of the Geronimo-Case, Borodin-Okounkov formula-- we find that \eqref{eq:correcchoice} is the only sensible  heuristic choice.  
\subsection[\qquad\qquad\qquad Asymptotics for the logarithmic derivative of the polynomials]{Asymptotics for the logarithmic derivative of the polynomials}
From the derivative formula \eqref{eq:dpdz}
\begin{equation}
    \frac{d \log P_N}{dz}=a_N(z)-b_N(z) \frac{P_N^*(z)}{P_N(z)}
\end{equation}
 or working with $Q_n(z)=P_n(z)e^{g(z)/2}$ instead,
 \begin{equation}
    \frac{d \log Q_N}{dz}=\tilde{a}_N(z)-b_N(z) \frac{P_N^*(z)}{P_N(z)}
\end{equation}
with $\tilde{a}_N(z)=a_N(z)+g'(z)/2$.

In the hydrodynamic limit with fixed $\lambda$, we expect
\begin{align}
    \frac{\tilde{a}_N(z)}{R}&\sim\alpha z+\frac{1+4\alpha u^2}{2}+\frac{\lambda+\lambda_c+4\alpha u^2}{z}+\frac{1+4\alpha u^2}{2z^2}+\frac{\alpha}{z^3}\\
   \frac{b_N(z)}{R}\frac{P_N^*(z)}{P_N(z)}&\sim \left(\alpha\left[2+\frac{2}{z^3}\right]+\left[1+2\alpha (2u^2-1)\right]\left[\frac{1}{z}+\frac{1}{z^2}\right]\right)\left(z-\sqrt{1-u^2}r(z)\right)
\end{align}
where we have used the Szeg\H{o} recursion to write 
\begin{equation}
\frac{P_n^*}{P_n}=\frac{z}{\alpha_n}-\frac{\rho_n}{\alpha_n}\frac{P_{n+1}}{P_n}, 
\end{equation}
\eqref{eq:ulambda}, and the ratio asymptotics \eqref{eq:correcchoice}. 
This means
\begin{equation}
    z-\sqrt{1-u^2}r(z)=\frac{z+1- \sqrt{(z+1)^2-4u^2 z}}{2}
\end{equation}
Putting everything together
\begin{equation}
    \frac{1}{R}\frac{d \log Q_N}{dz}=\frac{\sqrt{(1+z)^2-4 u^2 z} (z+1) \left(2 \alpha +4 \alpha  \left(u^2-1\right) z+2 \alpha  z^2+z\right)}{2 z^3}+\frac{\lambda }{z}
\end{equation}
or
\begin{equation}
\frac{z}{R} \frac{d \log Q_N}{dz}=\lambda +\left(2\alpha\left[z+\frac{1}{z}\right]+1-4\alpha[1-u^2]\right)\frac{z+1}{2\sqrt{z}}\sqrt{\frac{(z+1)^2}{z}-4u^2}.
\end{equation}
Setting $z=e^{\ci k}$ this is essentially the saddle point equation guessed from hydrodynamics. Indeed the rhs reads
\begin{equation}
    \lambda + \left(1+2\alpha[2\cos k+2(u^2-1)]\right)\sqrt{1+\cos k}\sqrt{\cos k+1-2u^2}
\end{equation}
or
\begin{equation}
    \lambda + \left(1+2\alpha[2\cos k+\cos k_c-1]\right)\sqrt{1+\cos k}\sqrt{\cos k-\cos k_c}
\end{equation}
where $k_c\in[0,\pi]$ and $\cos k_c=2u(\lambda)^2-1$ or $\cos \frac{k_c}{2}=u(\lambda)$, with $u(\lambda)$ given by (\ref{eq:ulambda}). This gives exactly the saddle point equation \eqref{eq:upsilondef} discussed in the main text. Note however that our derivation is a little bit sketchy, as we assume that the asymptotic result derived for $|z|>1$ extends to the portion of the unit circle $\{e^{\ci k},-k_c\leq k\leq k_c\}$.

\section{Free fermions calculations and numerical evaluations}
\label{app:freefermions}
\subsection{Exact two-point function}
Here we derive equation \eqref{eq:exactinverse}, which reads
\begin{equation}
    \braket{c_x^\dag c_{x'}}_{y,R}=\sum_{m,n=-L}^L T(R+y)_{x'm}T^{-1}(2R)_{mn} T(R-y)_{nx},
\end{equation}
where
\begin{align}\label{eq:Tmatrix_appendix}
    T(\tau)_{xx'}=\braket{0|c_x e^{\tau H}c_{x'}^\dag e^{-\tau H}|0}= \int_{-\pi}^{\pi} \frac{dk}{2\pi}e^{-\ci k(x-x')}e^{\tau \varepsilon(k)}
\end{align}
To do that, let us start from the definition
\begin{align}    \braket{c_x^\dag c_{x'}}_{y,R}=\frac{\braket{N|e^{(R-y)H}c_x^\dag c_{x'} e^{(R+y)H}|N}}{Z_N(R)}.
\end{align}
and use the anticommutation relations $c_x^\dag c_{x'}+c_{x'}c_x^\dag=\delta_{xx'}$ to obtain
	\begin{equation}
		\braket{c_x^\dag c_{x'}}_{y,R}=\delta_{x x'}-\frac{\bra{N}e^{(R-y)H}c_{x'}c_x^{\dag}e^{(R+y)H}\ket{N}}{\bra{N}e^{2RH}\ket{N}}.
	\end{equation}
Then  
\begin{equation}
 \begin{split}
	\bra{N}e^{(R-y)H}c_{x'}c_{x}^{\dag}e^{(R+y)H}\ket{N}
 &= \bra{0}c_{L}\cdots c_{-L} e^{(R-y)H}c_{x'}c_{x}^{\dag}e^{(R+y)H} c_{-L}^{\dag}\cdots c_{L}^{\dag} \ket{0}\\
 &=\bra{0} c_{L}(0)\cdots c_{-L}(0) c_{x'}(R-y) c_{x}^{\dag}(R-y) c^{\dag}_{-L}(2R)\cdots c^{\dag}_{L}(2R) \ket{0},
 \end{split}
	\end{equation}
where $c_x(\tau)=e^{\tau H}c_x e^{-\tau H}$. Now using Wick's theorem and the property
\begin{equation}
\bra{0}c_j(\tau_1)c^{\dag}_k(\tau_2)\ket{0}=\bra{0}c_j(0)c^{\dag}_k(\tau_2-\tau_1)\ket{0},
\end{equation}
we obtain
\begin{equation}
	\bra{N}e^{(R-y)H}c_{x'}c_{x}^{\dag}e^{(R-y)H}\ket{N}=\det M,
	\end{equation}
	with
	\begin{equation}
		M=
		\begin{pmatrix}
			\delta_{x x'} & T_{x'm} \\
			T_{nx} & T_{mn}
		\end{pmatrix},
	\end{equation}
where the matrix $T$ is defined in \eqref{eq:Tmatrix_appendix} and $m,n=-L,\dots, L$. A similar computation leads to
	\begin{equation}
		\bra{N}e^{2RH}\ket{N}= \det T.
	\end{equation}
Summarizing
	\begin{equation}
		\braket{c_{x}^\dag c_{x'}}_{y,R}= \delta_{x x'}- \frac{\det M}{\det T}
	\end{equation}
	and defining $\tilde{T}= \begin{pmatrix}
		1 & 0\\
		0 & T
	\end{pmatrix}$, we have
	\begin{equation}
		\begin{split}
			\frac{\det M}{\det T}=\frac{\det M}{\det \tilde{T}}= \det (M \tilde{T}^{-1})= \det \begin{pmatrix}
				\delta_{x x'} & T_{x'm} T^{-1}_{m n}\\
				T_{nx} & \mathbb{I}_{2L+1}
			\end{pmatrix}= \delta_{x x'} - \sum_{m,n=-L}^{L} T(R+y)_{x'+L,m} T^{-1}_{mn} T(R-y)_{n,x+L}.
		\end{split}
	\end{equation}
    This leads to
	\begin{equation}
		\braket{c_{x}^\dag c_{x'}}_{y,R}= \sum_{m,n=-L}^{L} T(R+y)_{x'm} T^{-1}(2R)_{mn} T(R-y)_{nx}.
	\end{equation}
\subsection{Two-point function and orthogonal polynomials}
Here we derive equation \eqref{eq:singleparticles} using a different method than in the main text. To start with let us write the matrix $T$ in \eqref{eq:Tmatrix_appendix} as
\begin{equation}
	\begin{split}
		T_{mn}	=\bra{0}c_m c^{\dag}_n(2R)\ket{0}&=\int_{-\pi}^{\pi} \frac{dk dq}{(2\pi)^2} e^{\ci qm} e^{-\ci kn} e^{2R \eps(k)} \bra{0}c(q)c^{\dag}(k)\ket{0}\\
		&= \int_{-\pi}^{\pi} \frac{dk}{2\pi} e^{\ci(m-n)k}e^{2R\eps(k)}=\oint_{\abs{z}=1} \frac{dz}{2\pi \ci z^{n-m-1}}e^{R(z+z^{-1}+\alpha(z^2+z^{-2}))}.
	\end{split}
\end{equation}  
So, $T_{mn}$ is a $(2L+1)\times(2L+1)$ Toeplitz matrix generated from the moments of $w(z):=e^{R(z+z^{-1}+\alpha(z^2+z^{-2}))}$. In particular, denoting as $\{P_k(z)\}_{k\geq0}$ such orthonormal polynomials of degree $j\geq 0$ corresponding to the weight $w(z)$, the inverse of the Toeplitz matrix $T_{mn}$ can be written as 
\begin{equation}\label{eq:inverseT}
	(T^{-1})_{mn}= \int_{-\pi}^{\pi} \frac{dk dq}{(2\pi)^2}e^{\ci((n+L)k-(m+L)q)}K_{2L}(e^{\ci q},e^{\ci k}),
\end{equation}
where $K_{2L}(e^{\ci q},e^{\ci k})$ is the Christoffel-Darboux kernel \eqref{eq:cd_def} and notice we shifted the indices $m,n \in\{-L,\ldots,L\}$ compared to the convention leading to  \eqref{eq:toeplitzinverse}.

Therefore the two-point function \eqref{eq:exactinverse} at $y=0$ is
\begin{equation}\label{eq:4integrals}
    \begin{split}
		\braket{c_x^\dag c_{x'}}_{0,R}= \sum_{m,n=-L}^{L} \int_{-\pi}^{\pi} \frac{\mathrm{d}k \mathrm{d}q \mathrm{d}\ell \mathrm{d}s}{(2\pi)^4} &e^{\ci k(x-m)}e^{\ci q(n-x')}e^{\ci((n+L)\ell -(m+L) s)}e^{R(\eps(k)+\eps(q))}K_{2L}(e^{\ci s},e^{\ci \ell}).
  \end{split}
\end{equation}
    We aim to expand the sum from $\sum_{m,n=-L}^{L}$ to $\sum_{m,n\in \mathbb{Z}}$, in order to retrieve the series representation of the Dirac delta given by $\delta(a-b)=\frac{1}{2\pi}\sum_{j\in\mathbb{Z} } e^{\ci(a-b)j}$. To achieve this, we notice that given a polynomial $f_n(z)=\sum_{k=0}^n a_k z^k$, for some coefficients $a_k$, the integral
	\begin{equation}
		\int_{-\pi}^{\pi} \frac{dq}{2\pi} e^{-\ci qj}f_n(e^{\ci q})= a_j,
	\end{equation}
	which means that it is a coefficients' extractor for the polynomial, so that
	\begin{equation}
		\sum_{j=0}^n 	\int_{-\pi}^{\pi} \frac{dq}{2\pi} e^{-\ci qj}f_n(e^{\ci q}) =\sum_{j\in \mathbb{Z}}\int_{-\pi}^{\pi} \frac{dq}{2\pi} e^{-\ci qj}f_n(e^{\ci q}),
	\end{equation}
	since $a_j=0$ for $j<0$ and $j>n$. Applying this consideration to \eqref{eq:4integrals} we arrive to 
    \begin{equation}
        \braket{c_x^\dag c_{x'}}_{0,R}=\int \frac{dk dq}{(2\pi)^2} e^{-\ci k(x+L)+\ci q(x'+L)}e^{R(\eps(k)+\eps(q))}K_{2L}(e^{\ci k},e^{\ci q}).
    \end{equation}
    Using the definition \eqref{eq:cd_def} and recalling $Q_l(z):=P_l(z) e^{\frac{R}{2}[z+1/z+\alpha(z^2+1/z^2)]}$ we obtain
   \begin{equation}
      \braket{c_x^\dag c_{x'}}_{0,R}=\sum_{l=0}^{N-1} \phi_l(x) \phi_l(x') 
   \end{equation}
   where $\phi_l(x):=\int \frac{dk}{2\pi} e^{-\ci k (L+x)} Q_l(e^{\ci k})$. This result is compatible with \eqref{eq:phik}.
\section{Some asymptotic calculations}
\label{app:besselfunction}
\subsection{Generalized Bessel function}
In the main text we introduced the generalized Bessel function
\begin{equation}
    J_n^{(\alpha)}(t)=\int_{-\pi}^{\pi} \frac{dk}{2\pi} e^{-\ci k n}e^{\ci t[\sin k-\alpha \sin 2k]}
\end{equation}
where $\alpha=0$ corresponds to the usual Bessel function $J_n^{(0)}(t)=J_n(t)$. Here we consider the case $\alpha\leq 1/8$. Various large $t,n$ limits can be accessed through standard saddle point considerations. To do that, rewrite
\begin{align}
    J_n^{(\alpha)}(t)= \int_{-\pi}^{\pi} \frac{dk}{2\pi} e^{\ci t \varphi(k)}
\end{align}
where $\varphi(k)$ depends implicitly on $n,t,\alpha$, and $t$ will be the large parameter. The saddle point equation $\varphi'(k)=0$ reads
\begin{align}
    \cos k-2\alpha \cos 2k=\frac{n}{t}
\end{align}
which has two real solutions $\pm k_s$ for $n/t\in (0,1-2\alpha)$. The leading behavior is then obtained by quadratic expansion of $\varphi(k)$ about the two saddle points.

Something peculiar happens when $n/t=1-2\alpha$. In this case the two saddle point coalesce at $k=0$, and the quadratic term vanishes. One then needs to expand to third order
\begin{align}
    \varphi(k)&=-k(1-2\alpha)+\sin k-\alpha \sin 2k\\\label{eq:cubicexpansion}
    &=(1-8\alpha)\frac{k^3}{6}+O(k^5).
\end{align}
This means
\begin{align}
    J_{t(1-2\alpha)+\sigma \left[t(1-8\alpha)/2\right]^{1/3}}^{(\alpha)}(t)&\simeq \int_{\mathbb{R}} \frac{dk}{2\pi}e^{-\ci k \sigma \left[t(1-8\alpha)/2\right]^{1/3}+\ci t (1-8\alpha)\frac{k^3}{6}}
\end{align}
and a simple change of variables shows
\begin{equation}\label{eq:Airyscaling}
  \lim_{t\to\infty} \; \left(\frac{t(1-8\alpha)}{2}\right)^{1/3} J_{t(1-2\alpha)+s\left[\frac{t(1-8\alpha)}{2}\right]^{1/3}}^{(\alpha)}(t) = \textrm{Ai}^{(3)}(s)
\end{equation}
which is the Airy function, see \eqref{eq:higherairyfunction} and comment below that. 
In the special case $\alpha=1/8$ this formula does not hold because the cubic term vanishes in \eqref{eq:cubicexpansion}, and one needs to expand up to order $5$ instead. We obtain
\begin{equation}\label{eq:higherAiryscaling}
    \lim_{t\to \infty}\; \left(\frac{t}{8}\right)^{1/5} J_{3t/4+s\left(\frac{t}{8}\right)^{1/5}}(t)\to \textrm{Ai}^{(5)}(s).
\end{equation}

\subsection{Partition function}
\label{app:scalingtofred}
To compute the partition $\tilde{Z}_N(R)$ as a Fredholm determinant, we will apply the Geronimo-Case, Borodin-Okounkov formula to the symbol $e^{g}$
\begin{equation}
    g(z)=z+\frac{1}{z}+\alpha\left(z^2+\frac{1}{z^2}\right)
\end{equation}
which implies $g^*(z)=z-\frac{1}{z}-\alpha\left(z^2-\frac{1}{z^2}\right)$. Using \eqref{eq:gc},\eqref{eq:gckern} and inserting the geometric series representation of $1/(1-w/z)$, we obtain the Fredholm determinant of the operator
\begin{align}
    \tilde{Z}_N(R)=\det_{\ell^2(\mathbb{N})} (1-K_N)
\end{align}
with kernel
\begin{equation}
    (K_N)_{ij}=\sum_{p\geq 0}J_{p+i+N+1}^{(\alpha)}(2R) J_{p+j+N+1}^{(\alpha)}(2R)
\end{equation}
Assume $0\leq \alpha<1/8$. 
In the refined limit $N=2R(1-2\alpha)+\sigma[(1-8\alpha)R]^{1/3}$ each $\textrm{Tr}\,\mathcal{K}^n$ becomes, introducing $\tilde{R}=R(1-8\alpha)$,
\begin{align}
    \textrm{Tr}_{\ell^2(\mathbb{N})}\,\mathcal{K}^n &=\sum_{i_1,\ldots,i_n=0}^\infty \mathcal{K}_{i_1i_2}\ldots \mathcal{K}_{i_ni_1}\\
    &=\sum_{i_1,\ldots,i_n=0}^\infty \sum_{p_1,\ldots,p_n=0}^\infty J_{p_1+i_1+N+1}^{(\alpha)}(2R) J_{p_1+i_2+N+1}^{(\alpha)}(2R)\ldots J_{p_n+i_n+N+1}^{(\alpha)}(2R) J_{p_n+i_1+N+1}^{(\alpha)}(2R)\\
    &\sim \tilde{R}^{-2n/3}\sum_{i_1,\ldots,i_n=0}^\infty \sum_{p_1,\ldots,p_n=0}^\infty \textrm{Ai}(\sigma+\frac{i_1+p_1}{\tilde{R}^{1/3}})\textrm{Ai}(\sigma+\frac{i_2+p_1}{\tilde{R}^{1/3}})\ldots \textrm{Ai}(\sigma+\frac{i_n+p_n}{\tilde{R}^{1/3}})\textrm{Ai}(\sigma+\frac{i_1+p_n}{\tilde{R}^{1/3}})\\
    &\sim \int_0^\infty dx_1\ldots dx_n \int_0^{\infty} d\lambda_1\ldots d\lambda_n \textrm{Ai}(\sigma+x_1+\lambda_1)\textrm{Ai}(\sigma+x_2+\lambda_1)\ldots \textrm{Ai}(\sigma+x_n+\lambda_n)\textrm{Ai}(\sigma+x_1+\lambda_n)\\
    &\sim \textrm{Tr}_{L^2([0,\infty))}\, \left[K_{\textrm{Ai},\sigma}^{(3)}\right]^n
\end{align}
where we have used \eqref{eq:Airyscaling}, and the kernel is
\begin{equation}
    K_{\textrm{Ai},\sigma}^{(3)}(x,y)=\int_\sigma^\infty d\lambda \textrm{Ai}(x+\lambda) \textrm{Ai}(y+\lambda)
\end{equation}
which is the Airy kernel. Hence
\begin{equation}
 \lim_{R\to\infty}  \tilde{Z}_{R(1-2\alpha)+s[(1-8\alpha)R]^{1/3}}(2R)= \det_{L^2([0,\infty))}(1-K_{\textrm{Ai},\sigma}).
\end{equation}
The rhs of the previous equation is the Tracy-Widom distribution \cite{TracyWidom1994}. A similar calculation based on \eqref{eq:higherAiryscaling} yields
\begin{equation}
     \lim_{R\to\infty}  \tilde{Z}_{3R/2+s[R/4]^{1/3}}(2R)=\det_{L^2([s,\infty))}(1-K_{\textrm{Ai},\sigma}^{(5)})
\end{equation}
The Fredholm determinant acts also on functions in $L^2(\mathbb{R}_+)$, but with higher order Airy kernel
\begin{equation}
K_{\textrm{Ai},\sigma}^{(5)}(x,y)=\int_0^\infty d\lambda \textrm{Ai}^{(5)}(x+\lambda) \textrm{Ai}^{(5)}(y+\lambda)
\end{equation}
and it defines a higher order Tracy-Widom distribution \cite{LeDoussalMajumdarSchehr2018}. There are an infinite number of them, and in general they have cumulative distribution function $F_{2n+1}(s)=\det_{L^2([s,\infty))}(1-K_{\textrm{Ai},\sigma}^{(2n+1)})$ where $n$ is a positive integer, as is also mentioned in the main text.

\bibliography{higher}

\end{document}